\documentclass{ar-astro2e}
\usepackage{ARAstroBib}
\usepackage{amsfonts}
\usepackage{amssymb}
\def\solm{M$_{\odot}\,$}
\def\soll{L$_{\odot}\,$}
\def\kms{km s$^{-1}$}
\def\hub{$\,$h$_{100}$$^{-1}\,$}
\def\etal {{et~al.\/}}
\def\deg{$^{\circ}\,$}
\def\solm{M$_{\odot}\,$}
\def\soll{L$_{\odot}\,$}
\def\kms{km s$^{-1}$}
\def\hub{$\,$h$_{100}$$^{-1}\,$}
\def\etal{{et~al.\/}}

\def\casgm20{CAS-G-M$_{20}\,$}
\def\m20{M$_{20}\,$}
\begin{document}
%\input epsf.tex    %<-If you need EPS figures to be
                   %  called in {figure} environment for PC
\input epsf.def   %<-If you need EPS figures to be
                   %  called in {figure} environment for Macintosh

\input psfig.sty

\jname{Annu. Rev. Astron. Astrophy.}
\jyear{2014}
\jvol{}
\ARinfo{1056-8700/97/0610-00}
\def\deg{$^{\circ}\,$}
\def\solm{M$_{\odot}\,$}
\def\soll{L$_{\odot}\,$}
\def\kms{km s$^{-1}$}
\def\hub{$\,$h$_{100}$$^{-1}\,$}
\def\etal   {{et~al.\/}}

\title{The Evolution of Galaxy Structure over Cosmic Time}

\markboth{Christopher J. Conselice}{The Structural Evolution of Galaxies}

\author{Christopher J. Conselice
\affiliation{Centre for Astronomy and Particle Theory \\ School of Physics and Astronomy \\ University of Nottingham, United Kingdom}}


\begin{abstract}

I present a comprehensive review of the evolution of galaxy structure in the universe
from the first galaxies we can currently observe at $z \sim 6$ 
down to galaxies we see in the
local universe.  I further address how these changes reveal galaxy formation 
processes that only galaxy structural analyses can provide.
This review is pedagogical and begins with a detailed discussion of the major methods
in which galaxies are studied morphologically and structurally.  This includes the 
well-established visual method for morphology; S{\'e}rsic fitting to measure galaxy 
sizes and surface brightness profile shapes; non-parametric structural methods including
the concentration ($C$), asymmetry ($A$), clumpiness ($S$)  (CAS) method, the Gini/M$_{20}$ 
parameters, as well as newer structural indices.   Included is a discussion of
how these structural indices measure fundamental properties of galaxies
such as their scale, star formation rate, and ongoing merger activity.    
Extensive observational results are shown demonstrating how broad galaxy morphologies 
and structures change with time up to $z \sim 3$, 
from small, compact and peculiar systems in the distant universe to
the formation of the Hubble sequence dominated by spirals and ellipticals
we find today.   This review further addresses how structural methods accurately 
identify galaxies in mergers, and allow measurements of  
the merger history out to $z \sim 3$.  The properties and evolution of
internal structures of galaxies are depicted,
such as bulges, disks, bars, and at $z > 1$ large star forming clumps.  The structure and
morphologies of host galaxies of active galactic nuclei and starbursts/sub-mm galaxies
are described, along with how morphological galaxy quenching occurs.  Furthermore, the
role of environment in producing structure in galaxies over cosmic time is treated.   
Alongside the evolution of general structure, I also delineate how galaxy sizes change
with time, with measured sizes up to a factor of 2-5 smaller at high redshift 
at a given stellar mass.   This review concludes with a discussion of how
the evolving trends in sizes, 
structures, and morphologies reveal the formation mechanisms behind
galaxies which provides a new and unique way to test theories of galaxy formation.

\end{abstract}

\maketitle

\section{INTRODUCTION}

Galaxy structure is one of the fundamental ways in which galaxy properties
are described and by which galaxy evolution is inferred.  There is a long
history of the development of this idea, which began with the earliest 
observations of galaxies, and continues up to the modern day as one of the 
major ways we study galaxies. This review  gives a detailed description 
of the progress made up to late-2013 
in using galaxy structure to understand galaxy formation and evolution.  It
is meant to be used as a primer for obtaining basic information from 
galaxy structures, including how they are measured and applied through cosmic
time.

The introduction to this review first gives an
outline of the basic events in the history of galaxy morphology and structure 
analyses, while the second part of the introduction describes how galaxy
structure fits into the general picture of galaxy formation.  I also
give a detailed
description of the goals of this review at the end of the introduction.

\subsection{Historical Background}

Galaxy morphology has a long history, one that even predates the time we
knew galaxies were extragalactic.  When objects which today we call galaxies 
were first observed what clearly distinguished them from stars
was their resolved structure.   Since this time, structure and morphology
has remained one of the most common ways galaxies 
are described and studied.  Initially this involved visual impressions of
galaxy forms. This has now been expanded to include quantitative methods 
to measure galaxy structures all the way back to the 
earliest galaxies we can currently see.

The first published descriptions of galaxy structure and morphology predates the 
telescopic era.  For example, the Andromeda nebula was described 
as a 'small cloud' 
by the Persian astronomer Abd al-Rahman al-Sufi in the 10th century, 
(Kepple \& Sanner 1998). The study of galaxies remained descriptive until the
late 20th century, although more and more detail was resolved as technology
improved.  As a result, for about 150 years the science of galaxies was
necessarily restricted to cataloging and general descriptions of structure, 
with notable achievements
by Messier and William and John Herschel who located galaxies
or `nebula' by their resolve structure as seen by eye.  Even before photography
revolutionized the study of galaxies some observers such as  William
Parsons, the 3rd Earl of Rosse, noted that the nebulae have a spiral 
morphology and first used this term to describe galaxies, most notably 
and famously in the case of M51.

It was however the advent of photography that astronomers could in earnest
begin to study the morphologies and structures of external galaxies.  The
most notable early schemes were developed by Wolf (1908), and
Lundmark (1926), among others.
This ultimately led to what is today called the Hubble classification which was
published in essentially its modern form in Hubble (1926), with the
final 'Hubble Tuning Fork' established in Hubble (1936) 
and Sandage (1961).  The basic Hubble sequence (Figure~1) consists of two 
main types of galaxies, ellipticals and spirals, with a further division 
of spirals into those with bars and those without bars.    Hubble, and the
astronomers who followed him, could classify most nearby bright galaxies in terms of
this system.

\begin{figure}%3	% Figure using psfig.sty
\centerline{\psfig{figure=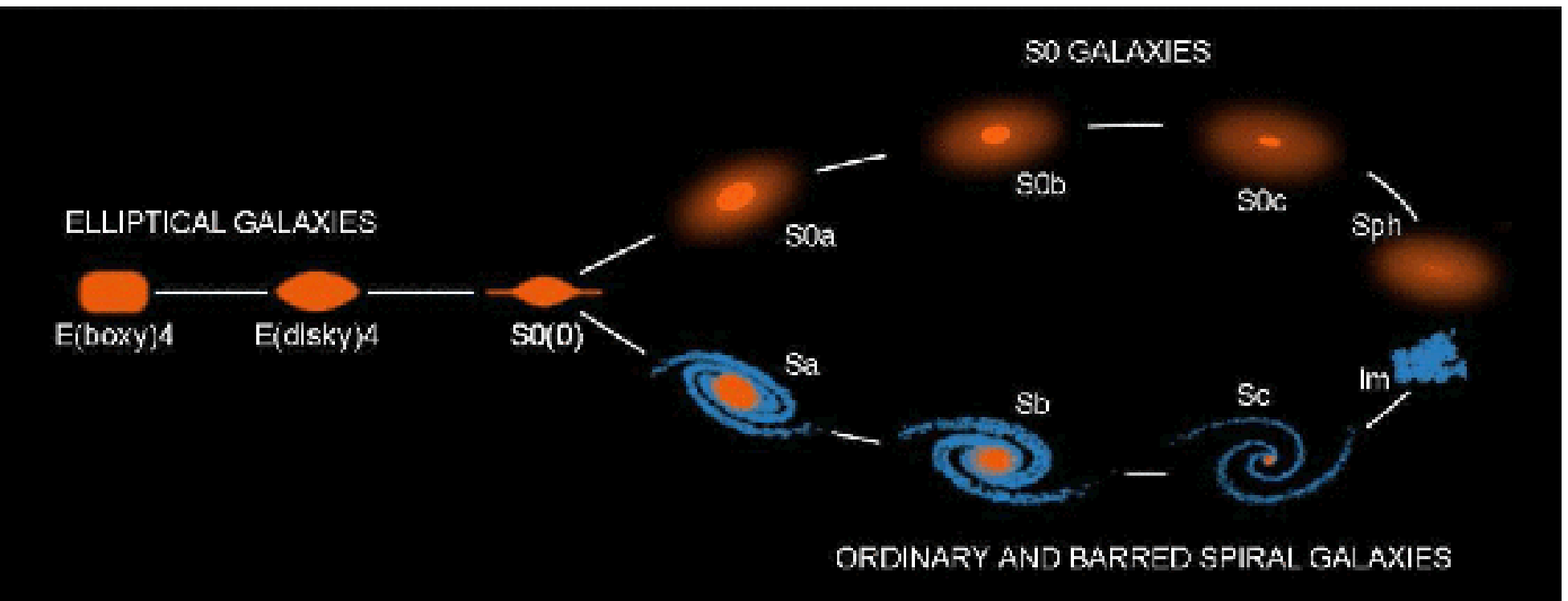,height=14pc}}
\caption{A modern form of the Hubble sequence showing the sequence
of ellipticals and S0s,  and the `tuning fork' in spirals.  The elliptical
sequence is determined by the overall shape of the galaxy, while 
spiral classifications are divided into different types (a-c) depending upon how 
wound-up spiral arms are, how large the bulge relative to the disk is, and how
smooth the spiral arms in the spirals arm.  The tuning-fork is the
differential between spirals with and without bars.   Also shown is the
extension of this sequence to dwarf spheroidal galaxies and irregular
galaxies, both of which are lower mass systems (Kormendy \& Bender
2012). }
\label{fig1}
\end{figure}

The development of morphological classification methods continued into the 
20th century, with newer methodologies based solely on visual 
impressions.   For example, de Vaucouleurs (1959) developed a revised 
version of the Hubble sequence which included criteria such as as bars, rings and
other internal features that were prominent on photographic
plates of galaxies. Likewise, van den Bergh (1960, 1976), and later
Elmegreen \& Elmegreen (1987) developed a system
to classify galaxies based on the form of spiral arms, and the apparent
clumpiness of the light in these arms.  

While it is important to classify galaxies visually, and all systems have 
some use, as all features should be explained by physics, it is not 
obvious which structural features of galaxies are 
fundamental to their formation history.   
Ultimately morphology and structure needs to prove 
to be useful for understanding galaxies, as there is now extensive
use of photometric and spectroscopic methods permitting measurements 
of perhaps more fundamental measures of 
stellar populations and dust/gas properties in galaxies. Along these lines, 
at roughly the same time as  progressively complicated classification systems were 
developed, astronomers such as Holmberg (1958)
established that physical properties of nearby galaxies correlate 
with morphology in a broad context.  Holmberg (1958) found that
ellipticals are typically massive and red,
and show little star formation, while spirals are less massive, bluer
and have evidence for ongoing star formation.   This quantitatively
expands into other physical parameters as well (e.g., Roberts 1963; Roberts
\& Haynes 1994; Conselice 2006a; Allen et al. 2006). It is also well known that
this segregation of morphology in the local universe provides an important
clue for understanding the physics of galaxy formation, especially as local
environment is found to strongly correlate with a galaxy's morphology (e.g., Dressler
1984; see \S 4.7).

A revolution in morphological and structurally studies came about
with the advent of photometric photometry, and especially the later
use of Charged Coupled Devices (CCD), which made detailed quantitative 
measurements of light distributions in galaxies possible.    The first major 
contribution from this type of work was by de Vaucouleurs (1948) who
used photometry to show
that the light profiles of what we would identify today as massive 
ellipticals all follow roughly the same fundamental light distribution, 
known as the de Vaucouleurs profile.  

This was later expanded by others, most notably S{\'e}rsic (1963), who demonstrated
that a more general form of light distributions matched galaxy light 
profiles with disks having exponential light profiles, while the light
distribution within massive ellipticals generally
following the de Vaucouleurs profile.  This has led to a huge industry in 
measuring the light profiles of galaxies in the nearby and distant 
universe which continues today (\S 2.2).

During the 1970s and 1980s the study of galaxy structure expanded to include
the decomposition of galaxy light into bulge and disk profiles (e.g., 
Kormendy 1977) as well as features such as bars, rings and lenses
(e.g., Kormendy 1979; de Vaucouleurs et al. 1993).  
The three dimensional structure of disk galaxies was investigated
(e.g., van der Kruit \& Searle 1982), as well as detailed studies
of bulges and disk in spiral systems (e.g., de Jong 1996; Peletier \&
Balcells 1996).  We also now know there is a great diversity in
elliptical galaxy internal structures (e.g., Kormendy et al. 2009).

Similar investigations demonstrated
that secular evolution within disks can provide an explanation for how
bars, rings and lenses can form  (e.g., Kormendy 1979; Combes \& Sanders 1981).
These effects, not driven by hierarchical galaxy formation, are also
likely responsible for the formation of pseudo-bulges and may drive
the formation of central massive black holes (e.g., Kormendy \& Kennicutt
2004; Sellwood 2013).

While there is a large amount of work done on the structures and morphologies
of galaxies in the nearby universe (e.g., see Kormendy et al. 2009;
Buta 2013), it is difficult
to investigate more than the very basics of structure and morphology when
studying distant galaxies.  This is due to the fact that current technology
does not allow us to resolve these distant galaxies in the same detail as
we can for closer systems.  As such, this review will concentrate on the
features and properties of galaxy structure which we can measure in
distant galaxies, and how this
reveals how galaxy evolution and formation occurs.  

The result of this is that one of the areas where galaxy structure and morphology 
has made its biggest impact is its ability to measure fundamental properties of
distant galaxies that we can compare with nearby galaxies to determine
evolution.   There are extensive methods for studying galaxy evolution
which galaxy structure analyses are becoming an essential aspect of, and
providing unique information on, the history and physics of galaxy 
assembly, which I detail 
in this review.    

\subsection{Galaxy Structure within the Context of Galaxy Formation}

We know that there is significant evolution in galaxies over time as the
stellar mass density of galaxies evolves rapidly at $1 < z < 3$, with
about half of all stellar mass formed by $z = 1$ (e.g., Bundy et al. 2005;
Mortlock et al. 2011).  We also know that there is a vast diversity of
star formation histories for individual galaxies, and that the integrated star
formation rate density in the universe's history peaks at $z \sim 2.5$,
and declines at higher and lower redshifts (e.g., Shapley 2011; Madau \& Dickinson
2014, this volume).  However, it 
is not clear from these observations what are/were the driving forces 
creating galaxies.

Theory offers several approaches for understanding how galaxies form which
detailed studies are starting to probe. We now believe that galaxy
formation can happen in a number of ways.  This includes: in-situ star formation in
a collapsed galaxy, major and minor mergers, and gas accretion from the
intergalactic medium.   Galaxy structure and
morphology are perhaps the best ways to trace these processes, as I discuss
in this review.

Another major question I address in this review is how do the structures and 
morphologies of galaxies change through cosmic time.  Major issues that
this topic allows us to address include: the formation history of the Hubble 
Sequence; whether galaxies form 'in-side-out' or 'out-side-in'?; how long
does a galaxy retain its morphology?; is morphology a invariant quantity
in a galaxy over a long cosmic time-span?, and furthermore what relative role does 
star formation and merging play in galaxy formation?

Galaxy structure and morphology has made a significant impact on these
questions largely due to the Hubble Space Telescope (HST) and its various
'Deep Field' campaigns starting in the mid-1990s, finding thousands of 
galaxies at redshifts $z >1$ within these images.   This is complemented
by extensive imaging and spectroscopy for nearby galaxies carried out by surveys
such as the Sloan Digital Sky Survey (SDSS) and the Millennium Galaxy
Catalog (e.g., Shen et al. 2003; De Propris et al. 2007).   Combining these surveys makes it
possible to study 
in detail the structures of distant galaxies, and to compare these with
structures at different redshifts.  This has led to a renaissance in the
analysis of galaxy structure, including parametric fitting using S{\'e}rsic
profiles, and the development of non-parametric measurements of galaxy
structure that have allowed us to use galaxy morphology/structure as a tool
for deciphering how galaxy assembly occurs over cosmic time. 

We are in fact now able to resolve galaxies back to redshifts
$z = 8$ with imaging from space, and recently as well with adaptive optics 
from the ground (e.g., Conselice \& Arnold 2009; Carrasco et al. 2010;
Akiyama et al. 2008). This reveals that galaxy structure is 
significantly different in the early universe from what it is today, 
and that there is a progression 
from the  highest redshifts, where galaxies are small, peculiar, and
undergoing high star formation rates to the relative quiescent galaxies that
we find in the nearby universe. How this change occurs, and what it implies 
for galaxy  evolution, is another focus of this review.   

Another ultimate goal is to describe the methods for measuring galaxy structure
and morphology for nearby galaxies up to the most distant ones we can see.  
I also discuss how galaxy structure correlates with physical properties
of galaxies, such as their star formation rate, merging and their overall
scale.  I then provide a description of the observed structural evolution of
galaxies, and a discussion of what this implies for the driving mechanisms
behind galaxy formation using the calibrated methods.

 The amount of information we have about the structures and 
properties of galaxies declines as one goes to higher redshift systems, 
and issues that arise due to observational bias must be dealt with.
I therefore also discuss systematics that can be addressed through imaging
simulations to determine the real evolution of the morphologies and
structures of galaxies. I finish this review with a discussion
of future uses of galaxy structure/morphology, including the potential
with the advent of JWST and Euclid.  

This review is structured as follows.  In \S 2, I describe the analysis
methods used for measuring the morphologies and structures of galaxies.
In \S 3 I describe how structures and morphologies reveal fundamental
galaxy properties and evolutionary processes, while \S 4 describes the 
observed evolution of
the structures of galaxies through cosmic time.  I finish this review 
with a description of how galaxy structure and evolution is becoming an 
important aspect for
understanding the underlying theory of galaxy formation and cosmology in
\S 5 and give a summary and future outlook in \S 6.

\section{Structural Measurement Methods}
 
In this section I describe the various ways in which galaxy structure 
is measured and quantified for comparisons across all redshifts.  
There are a great
diversity of nearby galaxy properties that cannot be examined at high
redshift, and this review only concentrates on that general features that
can be measured. This includes the traditional approach of using visual 
estimates to classify galaxies into morphological 
types, as well as quantitative methods.  Visual methods has had
a resurgence with the advent of Citizen Science projects such as Galaxy Zoo 
which provides online tools for non-scientists to classify over a 
million galaxies
(Lintott et al. 2011) as well as large Hubble Space Telescope projects 
such as CANDELS (e.g., Kocevski et al. 2011; 
Kartaltepe et al. 2014).  The bulk of this section however
describes the quantitative methods for
measuring galaxy structure, and the limitations to this approach.  
The interpretation of what these measurements imply are discussed
in \S 3 and \S 4.

\subsection{Visual Morphology}

The classic approach towards understanding the structures of galaxies
is through their apparent visual morphology.  
The major system of classification in use today has a development
through Hubble (1926), de Vaucouleurs (1959) and Sandage (1961, 1975) as
outlined briefly in the introduction. Modern reviews of galaxy
classification by eye into visual types can be found in Buta (2013).

When studying the morphologies of distant galaxies the visual classifications 
can only be placed into a few limited and well defined classes: 
spirals, ellipticals, and irregular/peculiars.  
The spirals can be further subdivided into spirals with or
without a bar.  In this review peculiars are interpreted as mergers
of two preexisting galaxies, while irregulars are lower mass galaxies
that contain a semi-random pattern of star formation, such as is
seen in Magellanic irregulars.  Typically these irregulars are
too faint to be seen at high redshifts and therefore are not
considered in this review in any detail.

Visual morphological classifications has been performed on
nearly all deep Hubble Space Telescope imaging starting from
its earliest days (e.g., Dressler et al. 1994;
van den Bergh et al. 1996).  This has continued with deeper and deeper HST
observations, including those that sample the rest-frame optical in the 
near-infrared
(e.g., Mortlock et al. 2013; Lee et al. 2013).  There are however
some limitations to how these classifications can be used at higher
redshifts, as it is not clear how the apparent morphology of a galaxy
will change due to redshift effects, rather than real evolution (\S 2.3.5). 

There is also the issue that galaxies which look 'elliptical' or 
'disky' do not have the same characteristics as systems with the same
morphologies seen nearby (\S 4.1).  It is clear that the properties of
distant galaxies that look elliptical and disky do not have the same 
physical properties as
systems with the corresponding morphology in the nearby universe.
Features such as sizes, light profiles, colors
and star formation rates differ within the same galaxy morphological type 
through time
(e.g., Conselice et al. 2011; Mortlock et al. 2013; Buitrago et al. 2013). 
Therefore, throughout this review a morphological type is only a visual 
determination of how a galaxy looks, and does not predispose to a certain 
local galaxy type or template, or to ascribe a certain formation history or scale.

\subsection{Parametric Measurements of Structure}

Historically one of the first ways in which galaxy structure was 
quantified
was through the use of integrated light profiles.  These profiles 
are simply
measured by taking the average intensity of a galaxy at a given radius, 
and
then determining how this intensity changes as a function of radius.
This was first described in detail by de Vaucouleurs (1948) who used the 
measurements of light from photometry at different apertures for
ellipticals and proposed
a fitting form.  A similar but more general form was found to 
better explain the surface brightness profiles by S{\'e}rsic (1963)
for different types of galaxies,

\begin{equation}
I(R) = I_{0} \times {\rm exp} (-b(n) \times R/R_{{\rm e}}^{1/n} - 1), 
\end{equation}

\noindent where the shape of the profile is described by the 
S{\'e}rsic index, $n$,
and the value of $b(n)$ is determined such that $R_{\rm e}$ is the
effective radius, containing half
of the light within the galaxy and is a function of the index $n$. The 
standard canonical benchmarks are that the de Vaucouleurs profile is 
given by $n=4$, and exponential disks by $n = 1$.  In 
principle, the values of $n$ and $R_{\rm e}$ are used as fundamental and
first order structural parameters of galaxies.  

The use of
the S{\'e}rsic profile to describe nearby galaxies is extensive (e.g., 
Kormendy et al. 2009), and
it has more recently been applied to distant galaxies, as we discuss
in \S 4.2.   For reviews on the use of resolve photometry through 
surface brightness profiles to study early type galaxies see 
e.g., Kormendy \& Djorgovski (1989), Allen et al. (2006), 
Simard et al. (2011)

Recently the fitting of galaxy two dimensional profiles with various forms
such as the S{\'e}rsic, exponential and de Vaucouleurs profile is done through the
GALFIT code by Peng et al. (2002), as well as GIM2D by Simard et al. (2011). 
These allow a simple and quick method for
measuring the light profiles and radii of many galaxies, providing data for
understanding the evolution of galaxy structure.  This allows for the 
measurements of different light components at high-z, although these 
codes
and other similar ones, have limitations such as a constant ellipticity 
assumption within a given component, but are sufficient for gross
measures of galaxy structure.

\subsection{Non-parametric Measurements of Structure}

Another more recent measurement technique involves the 
non-parametric method of  measuring galaxy light distributions.  Non-parametric methods of 
measuring galaxy structure began in the photographic era with attempts to
quantify the light concentration in galaxies by Morgan (1962), although 
extensive quantitative measures were not done
until the mid-1990s.

The development of methods to measure the light structures of galaxies 
began in earnest when the first deep images of distant galaxies were
obtained with the Hubble Space Telescope (Shade et al. 1995; 
Abraham et al. 1996) although their use
for low redshift measurements was also noted at about the same
time, although in terms of a physical property rather than a descriptive
quantity (e.g., Rix \& Zaritsky 1995; Conselice 1997; Bershady et al. 
2000; Conselice et al. 2000a,b).    These early papers show that
quantitative galaxy
structure correlates with other parameters, such as color and
peculiar features indicating mergers or galaxy interactions 
(e.g., Rix \& Zaritsky 1995;  Conselice 1997; Conselice et al. 2000a).

At present, the most common methods for measuring galaxy structure in a 
non-parametric way is through the CAS system (e.g., Conselice 2003; \S 2.3.1 - 2.3.3) 
and  through similar parameters (Takamiya 1999; Papovich
et al. 2003, 2005; Abraham et al. 2003; Lotz et al. 2004; Law 
et al. 2007; Freeman et al. 2013).   These parameters are designed 
to capture the major features of the 
underlying structures of these galaxies, but in a way that does not 
involve
assumptions about the underlying form, as is done with the S{\'e}rsic 
fitting
(\S 2.2).   These non-parametric parameters are also 
measurable out to high redshifts, making them ideal for deriving galaxy 
evolution over many epochs, as we discuss in \S 4.    

I give a brief description for how these parameters are measured.
Typically, as we discuss below, corrections must be applied 
to account for noise, and to use a reproducible
radius (e.g., Bershady et al. 2000; Conselice et al. 2000a).
This radius issue has been addressed in detail by e.g., 
Conselice et al. (2000a),
Bershady et al. (2000) and Graham et al. (2005).  The 
radius typically used in
these measures is the Petrosian radius,
which is  defined as the location where the ratio of surface 
brightness at a radius, $I(R)$, divided by the surface brightness within
the radius $<I(<R)>$,
reaches some value, which is denoted by $\eta(R)$ (Petrosian 1976). The 
value of $\eta$ changes
from $\eta(0) = 1$ at the center of a galaxy, down to $\eta(\infty) = 0$ 
when the light from the galaxy is zero at its outer 'edge'.   

This method of
measuring the radius is much less influenced by surface brightness dimming
than other methods such as using an isophotal radius, and is therefore
useful for measuring the same physical parts of galaxies at different 
redshifts (e.g., Petrosian 1976;  Bershady et al. 2000; Graham et al. 2005).  
The mathematical form for this radius is given by

\begin{equation}
\eta(R) = \frac{I(R)}{<I(<R)>},
\end{equation}

\noindent where most observables in non-parametric morphologies are measured
at a radius which corresponds to the location where $\eta(R) = 0.2$,  or
a relatively small multiplicative factor of this radius (often 1.5 times)
(e.g., Bershady et al. 2000; Conselice 2003; Lotz et al. 2004).

\begin{figure}%3	% Figure using psfig.sty
\centerline{\psfig{figure=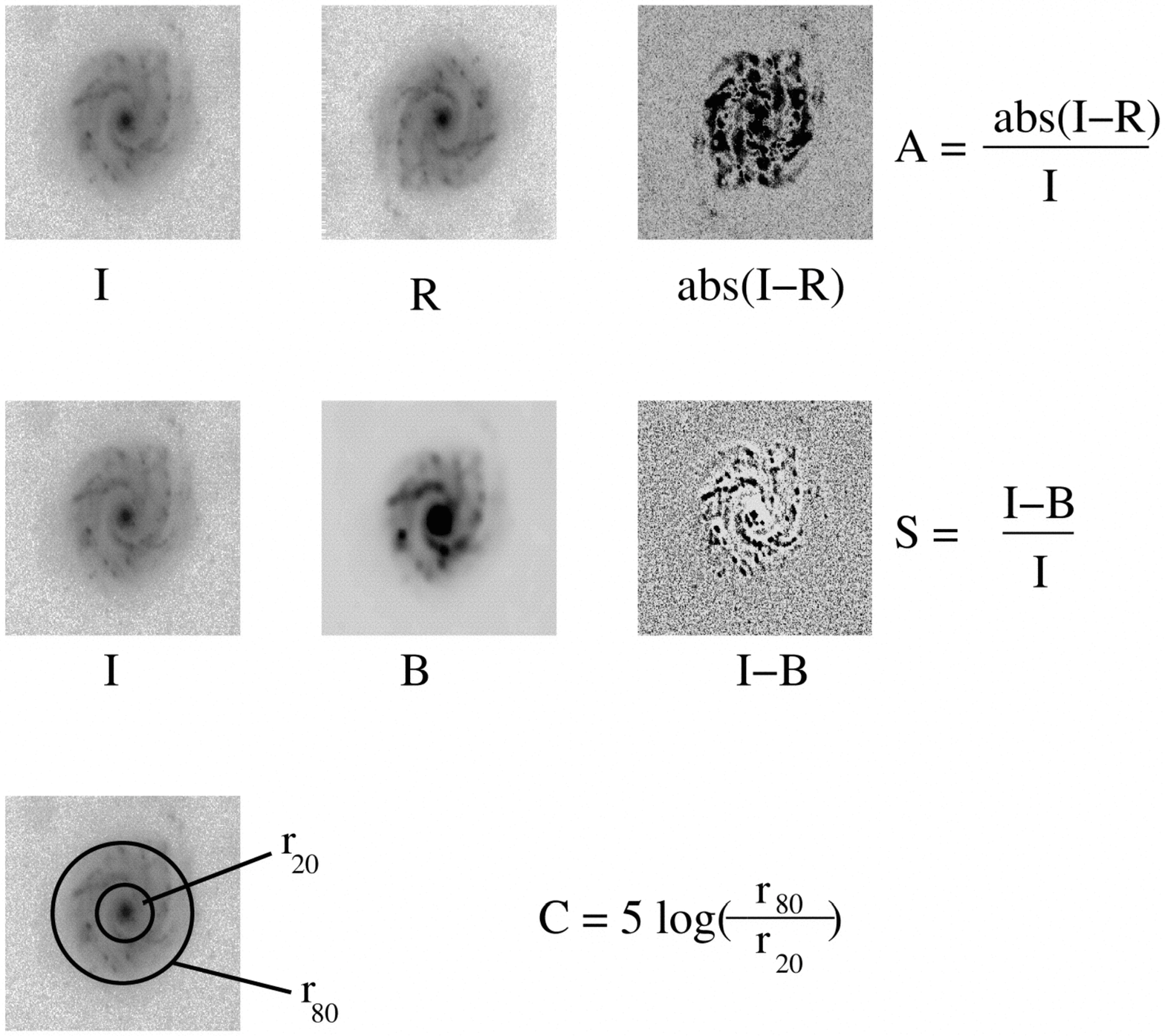,height=24pc}}
\caption{A graphical representation of how the concentration ($C$), asymmetry ($A$), clumpiness ($S$) are measured on an example nearby galaxy. Within the measurements for $A$ and $S$, the value '$I$' represents the original galaxy image, while '$R$' is this image rotated by 180 deg.  For the clumpiness $S$, '$B$' is the image after it has been smoothed (blurred) by the factor 0.3 $\times$ r($\eta = 0.2$). The details of these measurements can be found in Conselice et al. (2000a) for asymmetry, $A$, Bershady et al. (2000) for concentration, $C$, and
Conselice (2003) for the clumpiness index, $S$.}
\label{fig1}
\end{figure}

\subsubsection{Asymmetry Index}

One of the more commonly used indices is the asymmetry index($A$) 
which is a measure of how asymmetric a galaxy is after rotating 
along the line of sight center axis of the galaxy by 180 deg (Figure~2). 
It can be thought of as an indicator of what fraction of the light 
in a galaxy is in non-symmetric components.  

The basic formula for calculating the asymmetry index ($A$) is given by:

\begin{equation}
A = {\rm min} \left(\frac{\Sigma|I_{0}-I_{180}|}{\Sigma|I_{0}|}\right) - {\rm min} \left(\frac{\Sigma|B_{0}-B_{180}|}{\Sigma|I_{0}|}\right)
\end{equation}

\noindent Where $I_{0}$ represents is the original galaxy image, $I_{180}$ 
is the image
after rotating it from its center by 180\deg.  The measurement of the
asymmetry parameter however involves several steps beyond this simple
measure.  This includes carefully dealing 
with the background noise in the same way that the galaxy itself is by 
using a blank background area ($B_{0}$), and finding the location 
for the center of rotation.  The radius is usually defined as 
the Petrosian radius at which $\eta(R) = 0.2$, although once out to large
radius the measured parameters are remarkably stable.  

Operationally, the area $B_{0}$ is a blank part of the sky near the galaxy.
The center of rotation is not defined a priori, but is measured through an
iterative process whereby the value of the asymmetry is calculated at
the initial central guess (usually the geometric center or light centroid) and
then the asymmetry is calculated around this central guess using some 
fraction of a pixel difference.  This is repeated until a global minimum
is found (Conselice et al. 2000a).

Typical asymmetry 
values for nearby galaxies are discussed in Conselice (2003) with ellipticals
having values $A \sim 0.02\pm0.02$, while spiral galaxies are found in
the  range
from $A \sim 0.07-0.2$, while for Ultra-Luminous Infrared Galaxies (ULIRGs), which
are often mergers, the average is $A \sim 0.32 \pm 0.19$, and for
merging starbursts $A \sim 0.53 \pm 0.22$ (Conselice 2003).   Table~1 lists
the typical asymmetry and other CAS values (Conselice 2003). Quantitative 
structural values for the same galaxy can also differ significantly between wavelengths.
This is important for measuring these parameters at higher redshifts, 
where
often the rest-frame optical cannot be probed, an issue we discuss in
more detail in \S 2.3.5.

\begin{table}%
\def~{\hphantom{0}}
\caption{The average concentration ($C$), asymmetry ($A$), and clumpiness ($SS$) parameters for nearby galaxies as measured in the optical R-band 
(see Conselice 2003).}\label{tab1}
\begin{tabular}{lccc}%
\toprule
Galaxy Type & Concentration (R) & Asymmetry (R) & clumpiness (R)\\
\colrule
Ellipticals              & 4.4$\pm$0.3 & 0.02$\pm$0.02 & 0.00$\pm$0.04 \\
Early-type disks (Sa-Sb) & 3.9$\pm$0.5 & 0.07$\pm$0.04 & 0.08$\pm$0.08 \\
Late-type disks (Sc-Sd)  & 3.1$\pm$0.4 & 0.15$\pm$0.06 & 0.29$\pm$0.13 \\
Irregulars               & 2.9$\pm$0.3 & 0.17$\pm$0.10 & 0.40$\pm$0.20 \\
Edge-on Disks            & 3.7$\pm$0.6 & 0.17$\pm$0.11 & 0.45$\pm$0.20 \\
ULIRGs                   & 3.5$\pm$0.7 & 0.32$\pm$0.19 & 0.50$\pm$0.40 \\
Starbursts               & 2.7$\pm$0.2 & 0.53$\pm$0.22 & 0.74$\pm$0.25 \\
Dwarf Ellipticals        & 2.5$\pm$0.3 & 0.02$\pm$0.03 & 0.00$\pm$0.06 \\
\botrule
\end{tabular}
\end{table}

\subsubsection{Light Concentration}

The concentration of light is used as a method for quantifying how much
light is in the center of a galaxy as opposed to its outer parts.  It is
a very simple index in this regard, and it is similar to, and 
correlates strongly with, S{\'e}rsic $n$ values, which are also a 
measure of the light concentration in a galaxy.
There are many ways of measuring the concentration, including taking ratios of
radii which contain a certain fraction of light, as well as the ratio of the
amount of light at two given radii (e.g., Bershady et al. 2000; Graham
et al. 2005).  These radii are often defined by the total amount of light measured
within some Petrosian radius, often at the same location as used for the measuring
the asymmetry index.

The definition most commonly used is the ratio of two circular radii which 
contain a inner and outer fraction (20\% and 80\% or 30\% and 70\% are the most
common) ($r_{\rm inner}$, $r_{\rm outer}$) of the total galaxy flux (Figure~2),

\begin{equation}
C = 5 \times {\rm log} \left(\frac{r_{\rm outer}}{r_{\rm inner}}\right).
\end{equation}

\noindent    A higher value of $C$ indicates that a larger amount of light 
in a galaxy is contained within a central region.   The concentration index
however has to be measured very carefully, as different regions and radii
used can produce very different values that systematically do not reproduce
well when observed under degraded conditions (e.g., Graham et al. 2001;
Graham et al. 2005).

\subsubsection{Clumpiness}

The clumpiness (or smoothness) ($S$) parameter is used to 
describe 
the fraction of light in a galaxy which is contained in clumpy distributions.
Clumpy galaxies have a relatively large amount of
light at high spatial frequencies, 
whereas smooth systems, such as elliptical galaxies contain light at low 
spatial frequencies. Galaxies which are undergoing star formation tend to 
have very clumpy structures, and thus high $S$ values.  Clumpiness can be 
measured in a number of ways, the most common method used,
as described in Conselice (2003) is,

\begin{equation}
S = 10 \times \left[\left(\frac{\Sigma (I_{x,y}-I^{\sigma}_{x,y})}{\Sigma I_{x,y} }\right) - \left(\frac{\Sigma (B_{x,y}-B^{\sigma}_{x,y})}{\Sigma I_{x,y}}\right) \right],
\end{equation}

\noindent where, the original image $I_{x,y}$ is blurred to produce 
a secondary image,  $I^{\sigma}_{x,y}$ (Figure~2).  This blurred image is
then subtracted from the original image leaving a 
residual map, containing only high frequency structures in
the galaxy (Conselice 2003).  The size of the smoothing kernel $\sigma$ is
determined by the radius of the galaxy, and the value $\sigma = 0.2 \cdot 1.5
\times r(\eta = 0.2)$ gives the best signal for nearby systems 
(Conselice 2003).  Note that the centers of galaxies are removed when this procedure is carried 
out as they often contain unresolved high-spatial frequency light.

\begin{figure}%3	% Figure using psfig.sty
\centerline{\psfig{figure=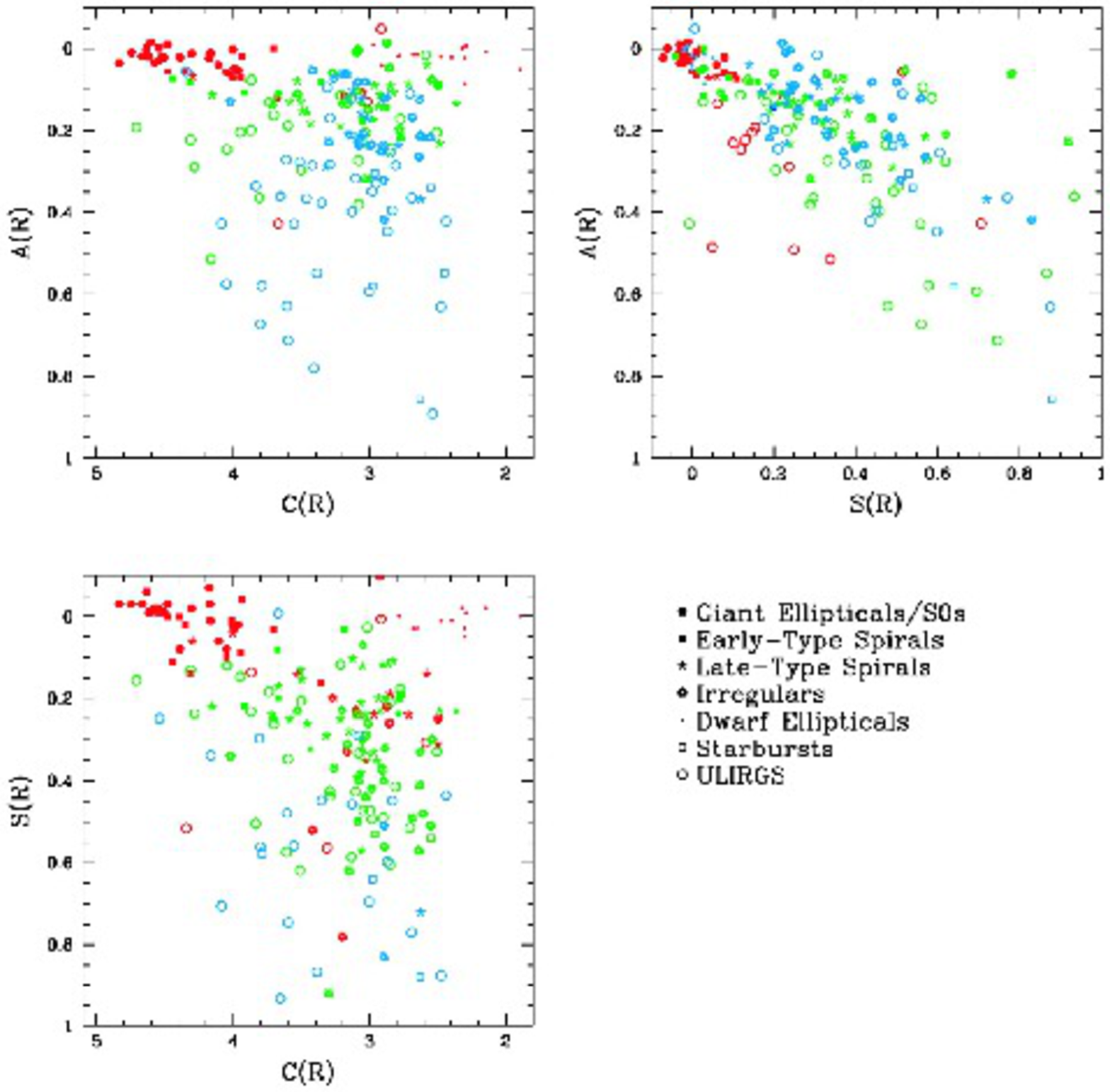,height=27pc}}
\caption{The different forms of the realizations of nearby galaxies of different morphologies and evolutionary states plotted together in terms of their CAS parameters. The top left panel shows the concentration and asymmetry indexes plotted with colored points that reflect the value of the clumpiness for each galaxy. Systems that have clumpiness values, $S < 0.1$ are colored red, galaxies with
values  $0.1 < S < 0.35$ are green, and systems with $S > 0.35$ are blue. In a similar way for the A-S diagram: red are for galaxies with for $C > 4$, green for systems with $3 < C < 4$, and blue points for $C < 3$. For the $S-C$ diagram: red is for systems with asymmetries $A < 0.1$, green have values $0.1 < A < 0.35$, and blue for $A > 0.35$.   When using these three morphological parameters all known nearby galaxy types can be distinctly separated and distinguished in structural space (Conselice 2003).}
\label{fig1}
\end{figure}

Figure~3 shows a diagram for how these three CAS parameters are measured
on a typical nearby spiral galaxy.
Furthermore, the CAS parameters can be combined together to create a 3-dimension space in which
different galaxy types can be classified.   For example, Figure~3 shows the  concentration 
vs. asymmetry vs. clumpiness diagram, demonstrating how these
parameters can be used to determine morphological types of galaxies in
the nearby universe in CAS space.

\subsubsection{Other Coefficients}

Another popular structural measurement system is the Gini/M$_{20}$ parameters
which are used in a similar way to the CAS parameters to find galaxies of
broad morphological types, especially galaxies undergoing mergers
(e.g., Abraham et al. 2003; Lotz et al. 2004).  Both of these
parameters measure the relative distribution of light within pixels,
and do not involve subtraction, as is used for the asymmetry and
clumpiness parameters, and therefore in principle may be less sensitive
to high levels of background noise (e.g., Lotz et al. 2004).

The Gini coefficient is a statistical tool originally used in economics to 
determine the distribution of wealth within a population, with higher values 
indicating a very unequal distribution (Gini of 1 would mean all wealth/light 
is in one person/pixel), while a lower value indicates it is distributed 
more evenly amongst the population (Gini of 0 would mean everyone/every pixel 
has an equal share).   The value of G is defined by the Lorentz curve 
of the galaxy's light distribution, which does not take into 
consideration spatial position.  

In the calculation of these parameters, each pixel is ordered by its
brightness and counted as part of the cumulative distribution (see
Lotz et al. 2004, 2008).     A galaxy in this case is considered
a system with $n$ pixels each with
a flux $f_{i}$ where $i$ ranges from 0 to $n$.   The Gini
coefficient is then measured by

\begin{equation}
G = \frac{1}{|\overline{f}|n(n-1)} \sum\limits_{i}^{n} (2i - n -1) |f_{i}|
\end{equation}

\noindent where $\overline{f}$ is the average pixel flux value.  The second
order moment parameter, \m20,
is similar to the concentration in that it  gives a value that indicates 
whether light is concentrated within an image. 
However a M$_{20}$ value denoting a high concentration (a very negative value) does
not imply a central concentration, as in principle the light could be concentrated
in any location in a galaxy.   The value of  \m20 is the moment of the 
fluxes of the brightest 20\% of light in a galaxy, which is
then normalized by the total light moment for all pixels (Lotz
et al. 2004, 2008).  The mathematical form for the M$_{20}$ index is

\begin{equation}
M_{20} = {\rm log} 10 \left(\frac{\sum\limits_{i}}{M_{\rm tot}} \right)\, {\rm while} \sum\limits{i} f_{i} < 0.2 f_{\rm tot}
\end{equation}

\noindent where the value of $M_{\rm tot}$ is 

$$M_{\rm tot} = \sum\limits_{i}^{n} M_{i} = \sum\limits_{i}^{n} f_{i} \left[(x_{i} - x_{c})^{2} + y_{i} - y_{c})^{2}\right]$$

\noindent where $x_{c}$ and $y_{c}$ is the center of galaxy, which in the case of M$_{20}$ this center
is defined as the location where the value of $M_{\rm tot}$ is minimized (Lotz et al 2004). 
The separation for nearby elliptical, spirals and ULIRGs is similar to that found by the
CAS parameters (see Lotz et al. 2004, 2008). 

Other popular parameters include the multiplicity index, $\Psi$, which is
a measure of the potential energy of a light distribution (e.g.,
Law et al. 2007).  Values of $\Psi$ range from 0 for systems that
are in the most compact forms, to those with values $\Psi > 10$ which
are often very irregular/peculiar (e.g., Law et al. 2012a).  Another
recent suite of parameters developed by Freeman et al. (2013) include
features that measure the 
multi-mode ($M$), intensity ($I$), and deviation ($D$) of a galaxy's
light profile with the intention to locate galaxy mergers.

\subsubsection{Redshift Effects on Structure}

One of the major issues with non-parametric structural indices is that they will
change for more distant galaxies, both due to any evolution but also due to 
distance effects, creating a smaller and fainter image of the same system.
This must be accounted for when using galaxy structure as a measure
of evolution (e.g., Conselice et al. 2000a; Conselice 2003; Lisker 2008).

There are several ways to deal with this issue, which is similar to how 
corrections for point spread functions in parametric fitting or
weak lensing analyses are done.  The most common correction method
for non-parametric parameters is to use image simulations.  These
simulations are such that nearby galaxies are reduced in resolution and surface
brightness to match the redshift at which the galaxy is to be simulated at.  These
new simulated images are then placed into a background appropriate for the
instrument and exposure time in which the simulation takes place (Conselice
2003).  The outline for how to do these simulations is provided in papers such as
Giavalisco et al. (1996) and Conselice (2003) amongst others.    

To give some idea of the difficultly in reproducing the morphologies and structures of 
galaxies, Figure~4 shows simulated nearby early-type spirals as to how they would appear
in WFC3 imaging data from the Hubble Ultra Deep
Field (Conselice et al. 2011a).  What can be clearly seen is that
it is difficult, and sometimes even impossible, to make out features of these
galaxies after they have been simulated.

Another issue when examining the structures of distant galaxies 
is that these systems will often be observed at bluer wavelengths than
what is typically observed with in the nearby universe due to the effects of redshift.
For example, pictures of galaxies at $z > 1.2$ taken with WFPC2 and ACS
are all imaged in the rest-frame ultraviolet.  Figure~5 shows what rest-frame
wavelength
various popular filters probe as a function of redshift.  This shows that
we must go to the near infrared to probe rest-frame optical light for
galaxies at $z > 1$.

It turns out that the qualitative and quantitative morphologies and structures
of galaxies can vary significantly between rest-frame ultraviolet and
rest-frame optical images (e.g., Meurer et al. 1995; Hibbard \& Vacca 1997;  Windhorst et al. 2002; Taylor-Mager et al. 2007) although
these morphologies are not significantly different for starbursting galaxies
with little dust at both low and high redshift (Dickinson 2000; Conselice
et al. 2000b).  While it is clear that the CAS method works better at 
distinguishing types at redder
wavelengths (e.g., Lanyon-Foster et al. 2012), its use has also expanded into 
image analyses with HI and dust-emission maps from Spitzer (e.g., Bendo
et al. 2007; Holwerda et al. 2011, 2013).

\begin{figure}%3	% Figure using psfig.sty
\centerline{\psfig{figure=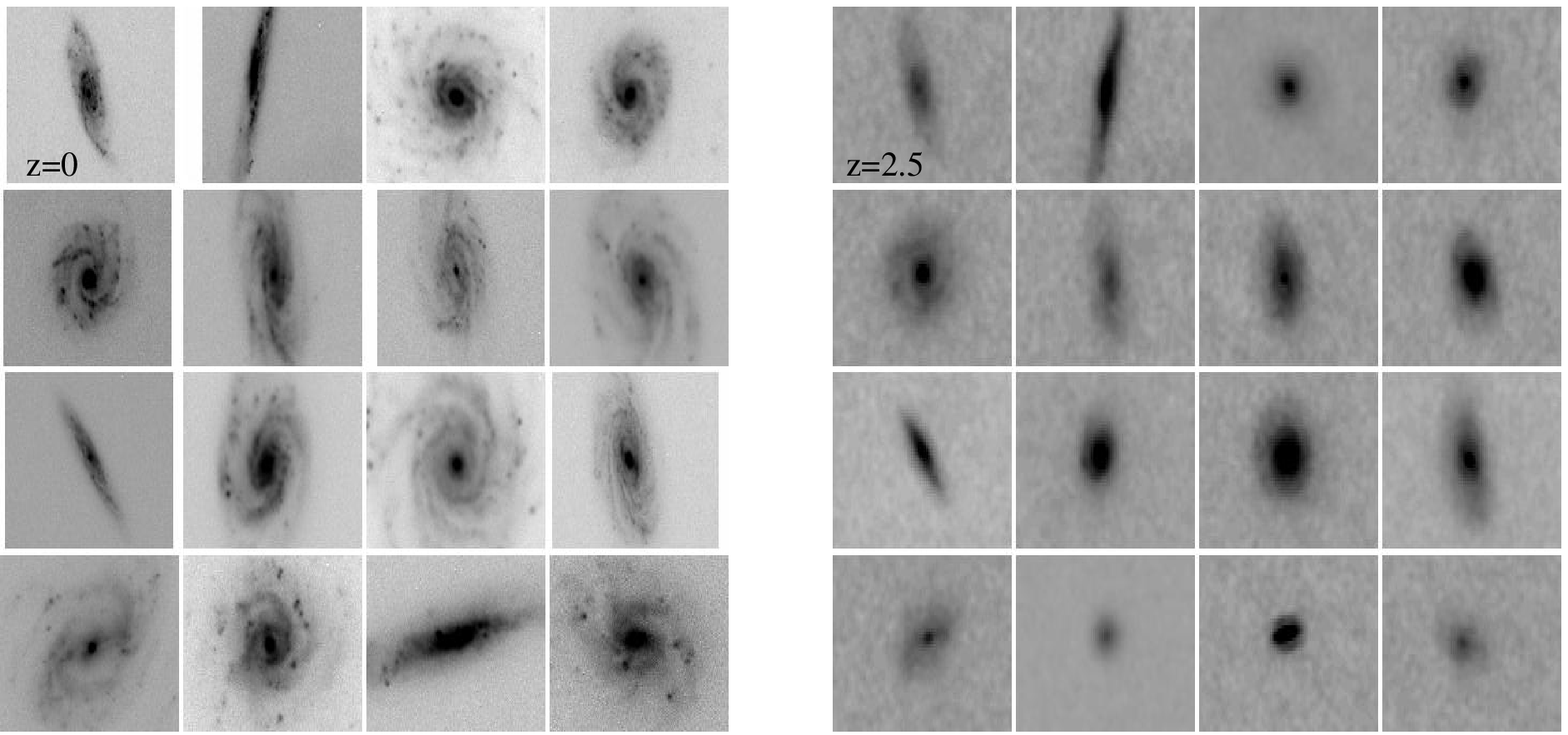,height=14pc}}
\caption{Nearby galaxies originally observed at $z \sim 0$ in the rest-frame B-band simulated to how they would appear at $z = 2.5$, also observed in the rest-frame B-band,  within the Hubble Ultra Deep Field WFC3 F160W (H) band. These systems are classified as late-type spirals (Sc and Sd) in the nearby universe but can appear very different when simulated to higher redshifts as can be seen here and when using quantitative measures.  The typical sizes of these galaxies are several kpc in effective radii, and are at a variety of distances (see Conselice et al. 2000a, Conselice et al. 2011a).  These changes in structure, both in apparent morphology
and in terms of the structural indices must be carefully considered before evolution is derived (e.g., Conselice et al. 2008; Mortlock et al. 2013).}
\label{Fig 1}
\end{figure}

\begin{figure}%3	% Figure using psfig.sty
\centerline{\psfig{figure=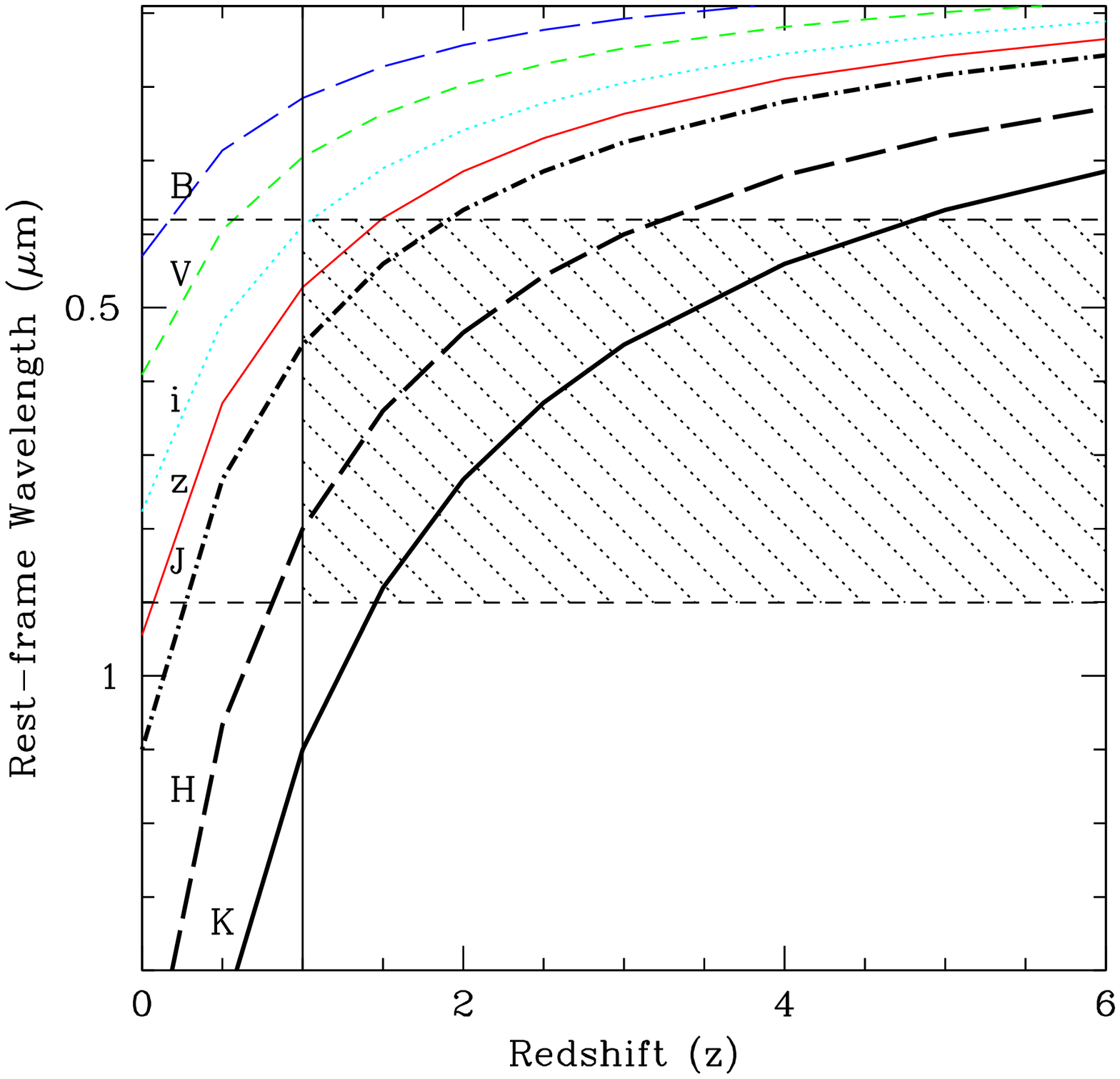,height=24pc}}
\caption{Plot showing the rest-frame wavelength probed by the most common
filters used to image distant galaxies as a function of redshift from $z \sim
0-6$.  The filters shown are the ACS B$_{450}$, V$_{550}$, i$_{775}$, z$_{950}$
filters, and the WFC3 J$_{110}$ and H$_{160}$ filters and a K-band filter 
centered at 2.2 $\mu$m.  The shaded area shows the region in which the 
rest-frame optical light of distant galaxies can be probed from between 
0.38 $\mu$m to 0.9 $\mu$m.  As can be seen, the H-band allows rest-frame 
light up to $z \sim 3$ to be imaged, while the K-band can extend this 
out to $z \sim 4.5$.}
\label{fig1}
\end{figure}

The process for accounting for the effects of image degradation  is to measure 
the morphological index of interest at $z = 0$, and then to remeasure the 
same values
at higher redshift after simulating.   For the morphological k-correction 
the approach has been to measure the parameter of interest at different wavelengths, and
to determine by interpolation the value at the rest-frame wavelength of
interest.  

Using the asymmetry index as an example, the final measure after correcting for redshift
effects is,

\begin{equation}
A_{\rm final} = (A_{\rm obs} + \delta A_{\rm SB-dim} + \delta A_{\rm k-corr}),
\end{equation}

% \times (1 - f_{\rm contam})

\noindent where $\delta A_{\rm k-corr}$ is the (usually negative) 
morphological k-correction, $\delta A_{\rm SB-dim}$ is the (positive) 
correction for images degradation effects. Other parameters can be
measured the same way. This is a necessary correction to examine 
the evolution of a selected population at the same effective
depth, resolution and rest-frame wavelength.  

\section{The Physical Nature of Galaxy Structure}

The ultimate goal in this review is to trace how structure evolves over
cosmic time, and using this as a method to decipher galaxy evolution.
As such, I present in this section work to date which describes 
how
physical information is derived from galaxy structural parameters.

\subsection{Star Formation and Galaxy Structure}

The star formation process within galaxies is critical, as galaxies
would not exist without stars in them.  Star formation is also
one of the major criteria for classification within the Hubble
sequence.   The effects of star formation
has also been used to classify spiral galaxies into various
classes (e.g., van den Bergh 1976; Elmegreen \& Elmegreen 1987).  

Star formation is an enormous topic with a large amount of work 
published (e.g., Kennicutt 1998; Kennicutt \& Evans 2012;
Madau \& Dickinson 2014), and we thus limit our discussion to how star
formation and galaxy structure are related.
The integrated star formation density evolution of galaxies in the 
universe has been studied in detail and is now
well characterized.  The integrated star formation rate 
increases from a low initial value at $z > 6$ to a peak at $z \sim 2$,
and thereafter declines.  At higher
redshifts $z > 1$ there is also a well defined relation between 
star formation rate and stellar mass, such that higher mass galaxies 
have a higher rate of star formation (e.g., Noeske et al. 2007; Bauer et al. 2011).

This is important as galaxies undergoing star formation  can
have very different morphologies and structures
from passive galaxies. Examples of this include: 
clumpy spiral arms,  knots of
star formation, central bright starbursts, etc.  This can be seen for 
example
when viewing local galaxies, whereby those with star formation
appear clumpier and more asymmetric than those without star formation.

Furthermore, there are also morphological k-correction effects, such
that star forming galaxies have a smaller difference in their morphology
between ultra-violet through optical and near-infrared light 
(e.g., Windhorst et al. 2002; Taylor-Mager et al. 2007). This generally reveals that
at shorter wavelengths the morphologies and quantitative structures
are tracing the distribution of star formation directly.  At optical
wavelengths longer than the Balmer break, we are sampling a
mixture of stars at different ages, with older ages dominating
the SEDs at longer wavelengths.   There are also very dusty galaxies
such as sub-mm sources and ULIRGs which can have a significant fraction of
their optical light absorbed (e.g., Calzetti et al. 2000).

It is not just apparently morphology which is affected by
star formation, but also the quantitative structure.  It is well
known that star forming galaxies without significant dust are
quite blue, but the effects of star formation can also be seen in their
structure.  Quantitative measurements of structure strongly correlates 
with the star formation rate within galaxies as measured by the 
correlation between the clumpiness index ($S$) and the 
H$\alpha$ equivalent width (Conselice 2003).  This is also seen
in more asymmetric and clumpy light distributions within
H$\alpha$ imaging of nearby galaxies, and when examining
the light distribution at 24 $\mu$m imaging using Spitzer Space
Telescope imaging (e.g., Bendo et al. 2007).  Conselice (2003)
calibrate how the clumpiness index can be used as a measure
of star formation, and Conselice et al. (2000a) show that
asymmetry values correlate strongly with (B-V) color for
nearby galaxies.

\subsection{Structure as a Merging Indicator}

One of the primary physical effects that can be seen in the structures
of galaxies is when two galaxies merge or interact with each other.   
When these dynamical events occur
the structures of these systems often become very peculiar and distorted,
especially when the merging galaxies contain a similar amount of mass
in a major merger\footnote{Note that a major merger throughout this review
is a merger where the ratio of the stellar masses of the progenitors are
1:3 or greater.  A minor merger is one with a mass ratio of less than
1:3.}  We have learned much about nearby galaxy mergers, such as
ULIRGs (e.g., Joseph \& Wright 1985; Sanders \& Mirabel 1996), as well
as through numerical simulations (e.g., Mihos \& Hernquist (1996) that 
have shown convincingly that peculiar galaxies are often mergers  
(Toomre \& Toomre 1972).  This demonstrates that there is a strong correlation
between structure and this fundamental galaxy formation process.

Early measurements of galaxy lopsidedness through Fourier decomposition of
structure, and by using the asymmetry parameter, measured on nearby 
galaxies found a correlation with the
presence of interacting or merging neighbors (e.g., Rix \& Zaritsky 1995; 
Conselice 1997; Reichard et al. 2008).
As such, galaxy structure is a powerful method for determining whether
a galaxy is undergoing a recent major merger.  This has been measured
in many ways, from using visual estimates of mergers based on peculiar
structures to more quantitative results.  

One automatic method for finding mergers is the CAS approach (Conselice
2003) where merging galaxies are those with a high asymmetry, which is
also higher than the value of the clumpiness.  The simple condition:

\begin{equation}
(A > 0.35)\, \&\, (A > S)
\end{equation}

\noindent accounts for a large fraction, but not all, of local 
galaxies which are mergers -- i.e., ULIRGs and starbursts in mergers (see
Figure~3).  
While the contamination from non-mergers is fairly low at a few percent,
the fraction of actual mergers which are identified is roughly
50\% (Conselice 2003).  This is largely due to the fact that galaxies
involved in the merger process are only quantitatively asymmetric for
about a third of the life-time of the merger (see \S 3.4). 

There is also the relationship found by Lotz et al. (2006) 
for locating major mergers using Gini/M$_{20}$ parameters is
given by,

\begin{equation}
G > -0.14\times M_{20} + 0.33.
\end{equation}

\noindent A more recent criteria developed by Freeman et al. (2013) uses
multi-mode ($M$), intensity ($I$), and deviation ($D$) statistics to 
quantify which galaxies are mergers. This study shows that a higher fraction
of real mergers can be found using these indices compared with CAS or
Gini/M$_{20}$.

One ultimate result of finding these mergers is that it allows
us to calculate the merger fraction within a population of galaxies.
The basic merger fraction ($f_{\rm m}$) is calculated as the number
of mergers selected within a given redshift bin and stellar
mass limit (or luminosity cut) ($N_{\rm m}$), divided by the total number of galaxies
within the same redshift and stellar mass selection ($N_{\rm T}$).   The merger 
fraction is thus defined as:

\begin{equation}
f_{\rm m}({\rm M_{*}}, z) = \frac{N_{\rm M}}{N_{\rm T}}.
\end{equation}

\noindent This merger fraction is 
also a function of stellar mass and redshift. The CAS mergers are
nearly all major mergers (Conselice 2003, 2006b; Lotz et al. 2008), 
whereas Gini/M$_{20}$ measures all types of mergers, both minor
and major.  

Furthermore for  structural samples we are calculating 
the merger fraction, 
as opposed to the galaxy merger fraction.   The difference is important when
comparing with pair studies where the two progenitors can be resolved. The
difference between these two is subtle, but important.  The merger fraction
considers a merger as having already happened, with the two galaxies that
have merged now counted as a single system. The galaxy merger fraction is
the fraction calculated when both of these galaxies merging are considered
as two separate galaxies, which they were before the final merger.  

The galaxy merger fraction ($f_{\rm gm}$) is thus the number of galaxies
merging, where a system which has already condensed into a single galaxy
is counted as two galaxies,
divided by the number of galaxies in the total sample. For small
merger fractions this ratio is about a factor of two larger
than the merger fraction which counts only the merger remnants 
(Conselice 2006b).
The equation to derive the galaxy merger fraction with observables
through morphology, with the assumption that every merging galaxy
has exactly two progenitor galaxies is given by,

\begin{equation}
f_{\rm gm}({\rm M_{*}}, z) = \frac{2 \times N_{\rm M}}{(N_{T} + N_{\rm M})} = \frac{2 \times f_{\rm m}}{(1+f_{\rm m})}.
\end{equation}

\noindent This relation does not hold if a merger occurs with more
than two galaxies (Conselice 2006b) although these are very rare
(de Propris et al. 2007).  The morphological measurement of the
nearby merger fraction gives values of $f_{\rm m}$ = 0.01 
(de Propris et al. 2007).  A discussion of the measurement of this at higher
redshifts is included \S 4.3.

\subsection{Galaxy Scale Properties and Galaxy Structure}

One of the interesting facts about galaxies is that many of their
characteristics can be explained by an underlying property, which
is likely its halo or total mass (Caon et al. 1993; Disney 
et al. 2008).   As an example, it was noted early on that galaxy light profile shapes of ellipticals
correlated strongly with the radius or magnitude of a galaxy
(e.g., Caon et al. 1993).  This implies that the scale or
mass of an elliptical galaxy has an influence on a galaxy's 
overall light profile and shape. 

This can also be seen in the detailed structures of galaxies. In
general it appears that on average galaxies with a higher degree of
central concentration have larger total or stellar masses.    
This is also seen in the concentration index, which
is another measure of the degree of light concentration, 
with more massive galaxies having a higher
value of concentration (e.g., Conselice 2003).  This concentration
also correlates with the fraction of light in bulge and disk
components. This relation is such that the more concentrated 
a galaxy is, the less likely it will contain a significant disk (e.g.,
Conselice 2003).  In fact, it is likely that it is 
the fraction of bulge light which drives this correlation, 
with more massive
systems more likely to have significant and concentrated
bulges.  

Concentration also separates galaxies with different
star formation histories in the local and high redshift universe.  In a
study using the Sloan Digital Sky Survey (SDSS) Strateva et al. (2001)
showed that non-star forming galaxies are more concentrated than star
forming blue systems.  This can also be seen with other overall galaxy
properties (e.g., Allen et al. 2006; Conselice 2006). The light
concentration for ellipticals also correlates with the mass of
the central massive black hole (e.g., Graham et al. 2002; 
Savorgnan et al. 2013).

\subsection{Numerical Simulations of Galaxy Structure}

Simulations of galaxy formation are critical for interpreting and
understanding the meaning of structural indices of galaxies.  
In fact, one of the
first computer simulations of galaxy formation 
by Toomre \& Toomre (1972) showed 
that the peculiar morphologies
of galaxies seen in e.g., the Arp (1966) atlas were due to systems
undergoing major mergers rather than some other cause.    Since then,
numerical simulations of galaxies have proven an effective method for
interpreting the structures and morphologies of galaxies in both the
local universe and at higher redshifts.  In many ways this approach
towards understanding galaxy morphology has just begun and promises
to be a powerful and effective approach for interpreting the meaning of
structure in the future.

\begin{figure}%3	% Figure using psfig.sty
\centerline{\psfig{figure=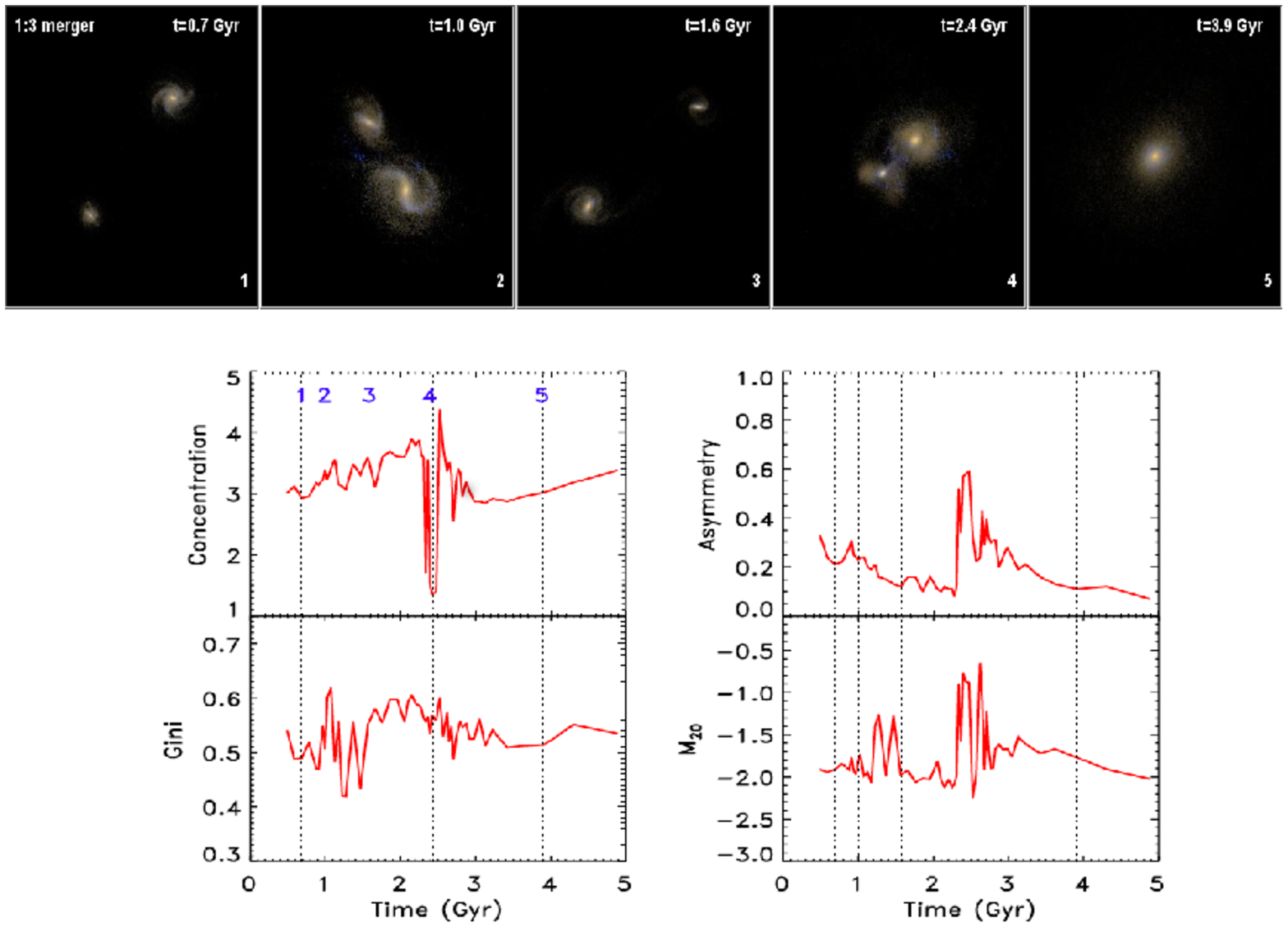,height=24pc}}
\caption{An N-body/hydrodynamical model from Lotz et al. (2008, 2010a,b) showing
two equal size disk galaxies merging as a function of time.  The numbers 
on the top of the realizations of this model show the various snap shots of
time through the simulation while the bottom panel shows the changes
in the asymmetry, Gini, concentration and M$_{20}$ values for this particular
simulation.  This demonstrates the changing form of quantitative indices during
a merger, and how these systems are only identifiable within the different
morphological systems as a merger at specific times. (courtesy Jennifer
Lotz)}
\label{fig1}
\end{figure}

Mihos \& Hernquist (1996) was one of the first papers to demonstrate 
in detail how the peculiar galaxies
seen in deep HST imaging were in fact due to the merger process using the
TREESPH hydrodynamical method  from Hernquist \& Katz  (1989).  
Interest in comparing simulated merger results 
with observables was largely in terms of 
starbursts and spectral energy distributions of galaxies 
(e.g., Barnes \& Hernquist 1991).    However, some attempts
were made even very early to use these merger N-body models as a method of
interpreting Hubble Space Telescope morphologies of galaxies (Mihos 1995).

Applying quantitative structural methods described in \S 2 to these 
numerical simulations of
structural evolution was first carried out by Conselice (2006b) who used the CAS
method to calculate the time-scale of the merger process 
on dark-matter particle simulations, finding a 
merger time-scale of $\sim 0.3-0.8$ Gyr for galaxies having 
a high enough asymmetry to be identified as an unambiguous merger (\S 3.2).  
This time-scale is critical for interpreting galaxy merger fractions
through cosmic time, as it allows us to convert merger fraction to 
merger rates,
and thus derive how mergers are driving galaxy formation.  
Using these results
Conselice (2006b) measure that a typical
galaxy undergoes 4.4$^{+1.6}_{-0.9}$ mergers from $z \sim 3$ to $z \sim 0$.

Conselice (2006b) also show that the time when a merger is asymmetric 
is distributed throughout the merger process, and is not located 
at one particular time.  Nor would the merging systems always be 
identifiable as
such when studied with the CAS parameters.  In addition 
to giving a robust time-scale, Conselice (2006b) also show that only a 
fraction (about a 1/3) of a merging galaxy's time-sequence would be found 
through CAS as a major merger. The time-scale derived 
is  also largely independent of the 
viewing angle of the merger and an asymmetry signal is only present 
within  major mergers with mass ratios of 1:3 or greater. 

The simulations used in Conselice (2006b) are however simple in that they do not
include the effects of star formation or dust, which are well know to produce dramatic
changes in morphology (e.g, Taylor-Mager et al. 2007; \S 3.4).  When star
formation and dust are added to simulations of galaxy structure the 
quantitative structural parameters measured are similar to those seen in nearby 
galaxies, and the measured structure correlates with other properties such
as color in the same way it does for nearby galaxies
(e.g., Lotz et al. 2008; Hambleton et al. 2011).

The paper Lotz et al. (2008) include the first measurements of 
CAS and Gini/M$_{20}$  parameters on numerical simulations that 
includes old and young stars, star formation, gas and dust.  Lotz 
et al. (2008) and later Lotz et al. (2010a,b) use 
GADGET/N-body/hydrodynamical simulations of galaxies when imaging 
the appearance of these galaxy mergers. 
Lotz et al. (2008) utilize disk galaxies of the 
same total mass, while Lotz et al. (2010a) investigate mergers with 
a variety of mass ratios, and Lotz et al (2010b) examine how the 
amount of cold gas mass in progenitor
galaxies influences morphology.    These 
simulation results are passed through the SUNRISE Monte-Carlo
radiative transfer code to produce, as realistically as possible, 
how galaxies would appear based on the simulation output.  

Lotz et al. (2008b, 2010a,b) further investigate the location in CAS and
Gini/M$_{20}$ parameter
space for mergers in different scenarios, and for different properties of
the merging galaxies.  They investigate
the time-scale for how long these simulated galaxies appear as a 'merger' 
based on where they fall in these non-parametric structural spaces 
(see Figure~6 for an example of these simulations). These 
papers also investigate how the dust, viewing angle, orbital 
parameters, gas properties, supernova  feedback and total mass alter 
the structural merger time-scale.  Lotz et 
al. (2008b; 2010a) find that most properties - the total mass, 
supernova feedback, 
viewing angle, and orbital properties of
mergers have very little influence on the derived time-scales. The 
mass ratio and gas mass fraction of the merging galaxies
affect the derived merger time-scales significantly however.

Mergers are identified within both CAS and Gini/M$_{20}$ at the first 
pass of the merger, as well as when the systems finally
coalesce to form a remnant (Lotz et al. 2008). However, 
merging galaxies are
not found in the merger area of the non-parametric structural parameters 
for the entire merger.
This however allows the time-scales for structural mergers to be calculated. 
Lotz et al. (2008, 2010a) find that the asymmetry time-scales for gas-rich 
major mergers are 0.2-0.4 Gyr, and 0.06 Gyr for minor mergers 
(Lotz et al. 2010a).   
The Gini/M$_{20}$ time scales are $\tau_{\rm m} = 0.2-0.4$ Gyr.   
These are relatively
quick time-scales, and thus suggests that the observed merger fraction 
converts to a high
merger rate.

This is similar, but not exactly the same as what is calculated for merger time
scales from dynamical friction for merging objects to have a separation change from $r_{i}$ to $r_{f}$.  The dynamical friction time-scale is given by, $t_{fric}$,

\begin{equation}
t_{\rm fric} = 0.0014 {\rm Gyr} \left(r_{i}^{2} - r_{f}^{2}\right) \left(\frac{v_{c}}{100 {\rm km s^{-1}}}\right) \left(\frac{10^{10} {\rm M_{\odot}}}{M}\right)
\end{equation}

\noindent where $v_{c}$ is the relative velocity between the two merging 
galaxies at a given time, $M$ is the mean accreted mass, and the Coulomb logarithm 
ln $\Lambda = 2$  (Dubinski et al. 1999).   Dynamical friction calculations
such as these have dominated the calculation of galaxy time-scales up until
simulations of mergers reveled more subtle results, although the blunt
calculations from eq. (13) are often a good rough estimate for merger 
time-scales, giving values of $\sim 0.5$ Gyr for equal mass mergers.

Lotz et al. (2008b, 2010a) also find that the asymmetry index is sensitive to major
mergers of ratios of 1:4 or less, while the Gini/M$_{20}$ is sensitive
for mergers down to 1:9 -- thus probing more minor mergers.   Lotz
et al. (2010b) however find that very gas rich galaxies, such as those
seen in high redshift may have longer time-scales for merging with gas
rich progenitors which are likely to be more common at higher redshifts.
This would provide a 'merger' asymmetry signal for more minor mergers as
long as they were more gas rich.  However it is clear that massive
galaxies with M$_{0} > 10^{10}$ \solm, where most measurements have been 
made to date at $z < 3$, have a low gas mass fraction (e.g., Erb
et al. 2006; Mannuci et al. 2010; Conselice et al. 2013)

\section{Measurements of Galaxy Structural Evolution}

The above sections describe how we can measure the
structures and morphologies of galaxies through various approaches, and
the meaning of this structure.  In this section I discuss
how these measurements have been applied to galaxies at all redshifts
to decipher how evolution is occurring within the galaxy population. 

When galaxies were first found in the distant universe, they were 
not resolved enough to study their structures and morphologies, and the 
evolution of galaxies was
observationally driven by number counting and colors (e.g., Koo \& Kron
1992), with the 'faint blue galaxy excess' problem at faint magnitudes
dominating the field for twenty years, until redshifts
for these systems became available (e.g.,  Ellis 1997).

The problem of galaxy evolution and formation is a large one, and
this review does not focus on this question.  However, galaxy
structure and morphology reveals information
that cannot be provided by other methods.  I do however give a brief 
overview here of the important questions in understanding galaxy evolution.
For a general review of galaxy properties at $z > 2$ see Shapley (2011)
for an observational perspective and Silk \& Mamon (2012) for
a theoretical one.  For
nearby galaxy's a few recent relevant reviews are 
Blanton \& Moustakas (2009), van der Kruit \& Freeman (2011),  and
Conroy (2013).

Galaxies are now studied up to redshifts $z \sim 7-10$, although at the 
highest redshifts less information is available.  The most common 
measures for these
distant galaxies are colors, stellar masses, star formation
rates, sizes and basic structures.  From this we know that the volume integrated
star formation rate increases with time from these ultra-high redshifts until
around $z \sim 2$ when the star formation rate begins to decline (see Madau
\& Dickinson 2014, this volume).  Stellar mass
measurements roughly agree with this picture, such that about half of all stellar
mass is formed by $z \sim 1$ (e.g., Mortlock et al. 2011).  Galaxies are also much
bluer in the past (Finkelstein et al. 2012), with some debate and uncertainty 
concerning the star formation
history of individual galaxies and the relevant role and commonality 
of very old and/or very dusty galaxies at redshifts $z > 1$.

What is largely unavailable from examining stellar masses, stellar populations and
star formation histories is how these galaxies assembled. Clearly galaxies are fed
gas, or have very large gas reservoirs in them to sustain and produce star formation.
How this gas gets into galaxies is a fundamental question, as is the relative role
of mergers vs. star formation in forming galaxies.  Since the number of massive
galaxies at high redshifts is a factor of ten or so less than those today, clearly
much evolution and formation in these systems has occurred.

Galaxy structure provides a way to examine this problem, as it permits us to determine
which modes of galaxy formation are active within a galaxy.  The first and by far
the most common method is to study the merger history through the techniques 
described in \S 3.2.  Another method is to simply examine the visual morphologies of
galaxies to determine when the Hubble sequence is in place, and after combining with
color, size and star formation rates, to determine when spirals and ellipticals 
are roughly in 
their current form.    In particular the examination of the sizes of galaxies has 
provided a very puzzling evolution such that galaxies of similar masses are up 
to a factor of 2-5 times smaller than corresponding galaxies seen today 
(e.g., Buitrago et al. 2008; Cassata et al. 2013).  

Furthermore, resolved imaging allows us to study the formation history of
individual components of  galaxies, such as disks, bulges, spiral arms, clumps 
of star formation, etc which reveals formation information not available when
examining the galaxy as a whole.  This section, which is the heart of this review
article, provides the current observational evidence for morphological and
structural evolution and what it implies for galaxy formation. 

\subsection{Observed Evolution of the Hubble Sequence}

The first science result I discuss is how the visual Hubble
sequence evolves throughout the universe. This can simply be
restated as measuring the number density and relative fractions of
galaxy types at a given selection, which are classified as ellipticals, spirals
and peculiars. As mentioned earlier, we do not consider irregulars as these
are typically lower mass galaxies that are not detected at high redshift
due to their faintness.   It is also largely impossible to use finer classifications,
such as Sa or Sb, on distant galaxies, as the resolution is not good 
enough, even with Hubble Space Telescope imaging, to resolve this type of detail.
More distant galaxies also appear to be quite different from Hubble types,
making this type of detailed morphologies unnecessary and 
undesirable (e.g., Conselice 2005).
  
This relates to a fundamental question that has been asked since galaxies were
 discovered, which is whether or not a galaxy retains its morphology over a long 
period of cosmic time. Otherwise, if morphological transformations do occur, 
what processes drive this (internal or external), and how often does  a galaxy
change its morphology?

\begin{figure}%3	% Figure using psfig.sty
\centerline{\psfig{figure=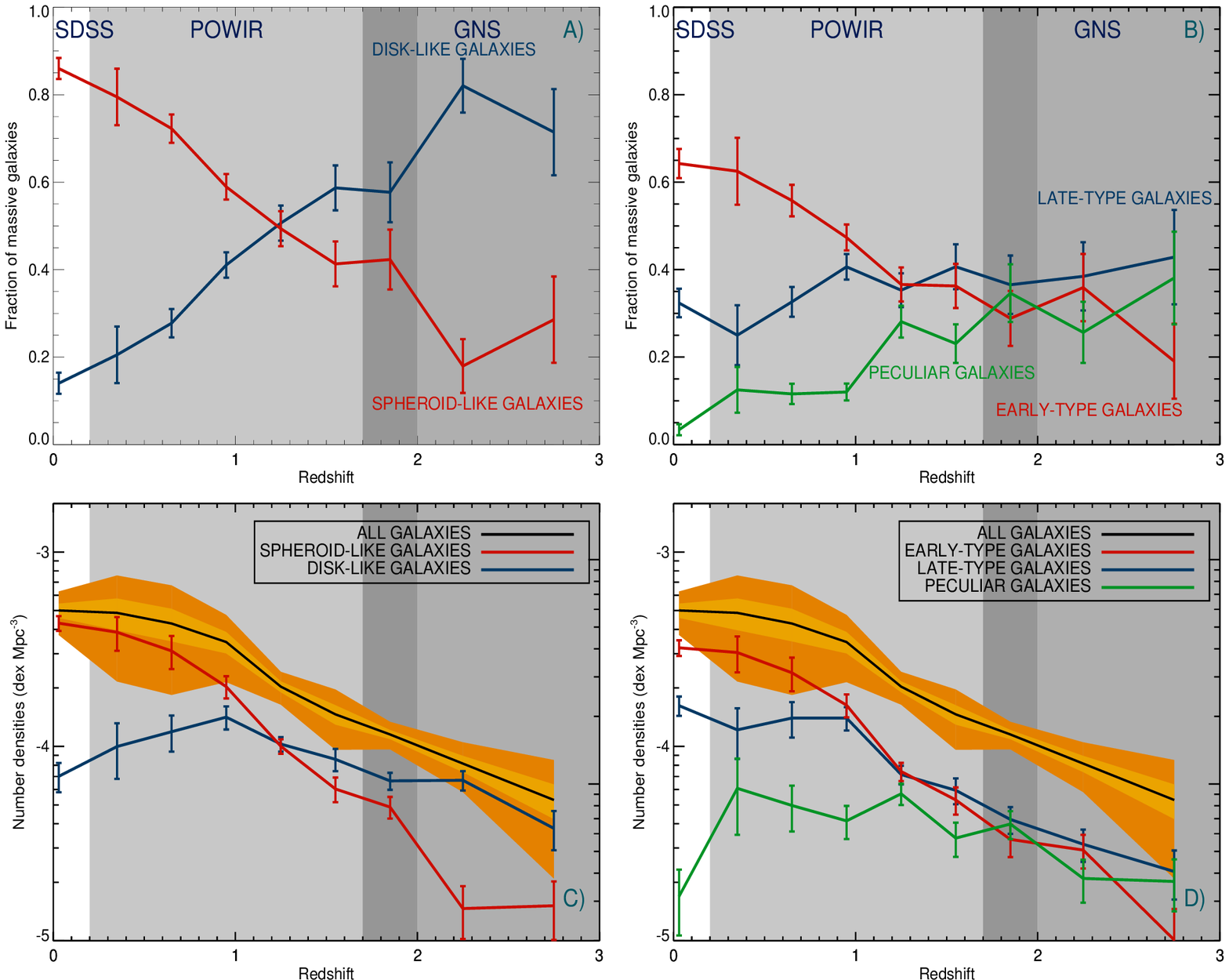,height=27pc}}
\caption{The evolution of apparent morphology, and S{\'e}rsic index
based classifications for massive galaxies with
M$_{*} > 10^{11}$ \solm from Buitrago et al. (2013).  Shown are
the morphologies and structures derived from three different surveys -- SDSS 
for nearby galaxies, POWIR survey (Conselice et al. 2007) for systems
up to $z \sim 2$, and the GOODS NICMOS Survey for systems for systems at $z \sim
1.8 - 3$.  In this plot, disk-like galaxies are those with
S{\'e}rsic indices $n < 2.5$, and spheroid-like galaxies are for those with
$n > 2.5$.  The right panel shows the morphological evolution as judged from
visual estimates.  Plotted here is both the fraction of types, as well
as the number density evolution.  The orange shading gives the total number
density of all galaxies as a function of redshift up to $z \sim 3$.  As can be seen,
there is a gradual transition from galaxies that appear peculiar and
'disk-like' in their S{\'e}rsic indices at high redshifts $z > 1.5$ which
gradually transform into early types today.   }
\label{fig1}
\end{figure}

One of the first observations noted when examining the first deep HST images was that 
many of the fainter galaxies were peculiar (e.g.,
Driver et al. 1995; Abraham et al. 1996; Schade et al. 1995).  
These early studies were limited to examining galaxy number counts, 
as no redshifts were 
known for these faint galaxies.  In addition to a faint blue excess, 
it was clear that there was also a peculiar excess, and often for the
faintest galaxies.  This is where the field remained until redshifts for
a significant number of these peculiar galaxies were obtained.

\begin{figure}%3	% Figure using psfig.sty
\centerline{\psfig{figure=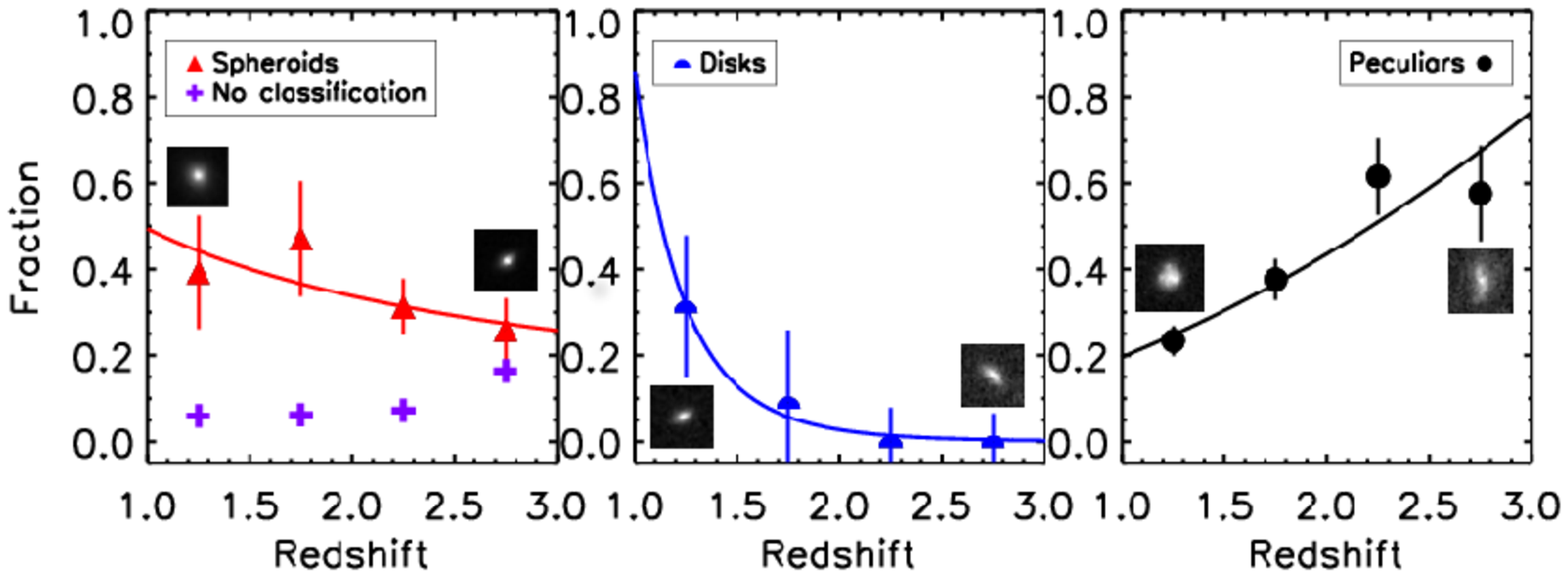,height=16pc}}
\caption{The latest version of the evolution of the Hubble sequence with redshift
for galaxies with M$_{*} > 10^{10}$ \solm.  These classifications are corrected
for image degradation such that misclassification due to distance are accounted
for within these fractions (Mortlock et al. 2013). Examples of images of galaxies in each of
these bins is shown in the observed $H_{160}$-band or rest-frame optical. 
Further analysis shows that there is a downsizing trend such that the most 
massive galaxies form into Hubble sequence galaxies earlier than lower mass 
galaxies (Mortlock et al. 2013). }
\label{ }
\end{figure}

The field of high redshift studies changed dramatically in 1995-1997 with the
advent of the both the Hubble Deep Fields (HDF) (Williams et al. 1996; 
Ferguson et al. 2001) and the discovery of a significant population of
high-redshift galaxies that could be discovered by the Lyman-Break technique
(Steidel et al. 1996; Shapley 2011) now referred to as Lyman-Break Galaxies (LBGs).  
The HDF, and 
later significant campaigns to obtain very deep Hubble 
imaging, such as the Hubble Deep Field South (Williams et al. 2000), the
Great Observatories Origins Deep Survey (GOODS) (Giavalisco et al. 2004),
the Hubble Ultra Deep Field (UDF) (Beckwith et al. 2006), the 
COSMOS field 
(Scoville et al. 2007), the Extended Groth Strip survey (EGS) (Davis et al. 2007),
and most recently the CANDELS survey (Grogin et al. 2011; Koekemoer et al. 2011), 
have revolutionized the field of galaxy formation studies and in particular 
the study of morphologies.

\begin{figure}%3	% Figure using psfig.sty
\centerline{\psfig{figure=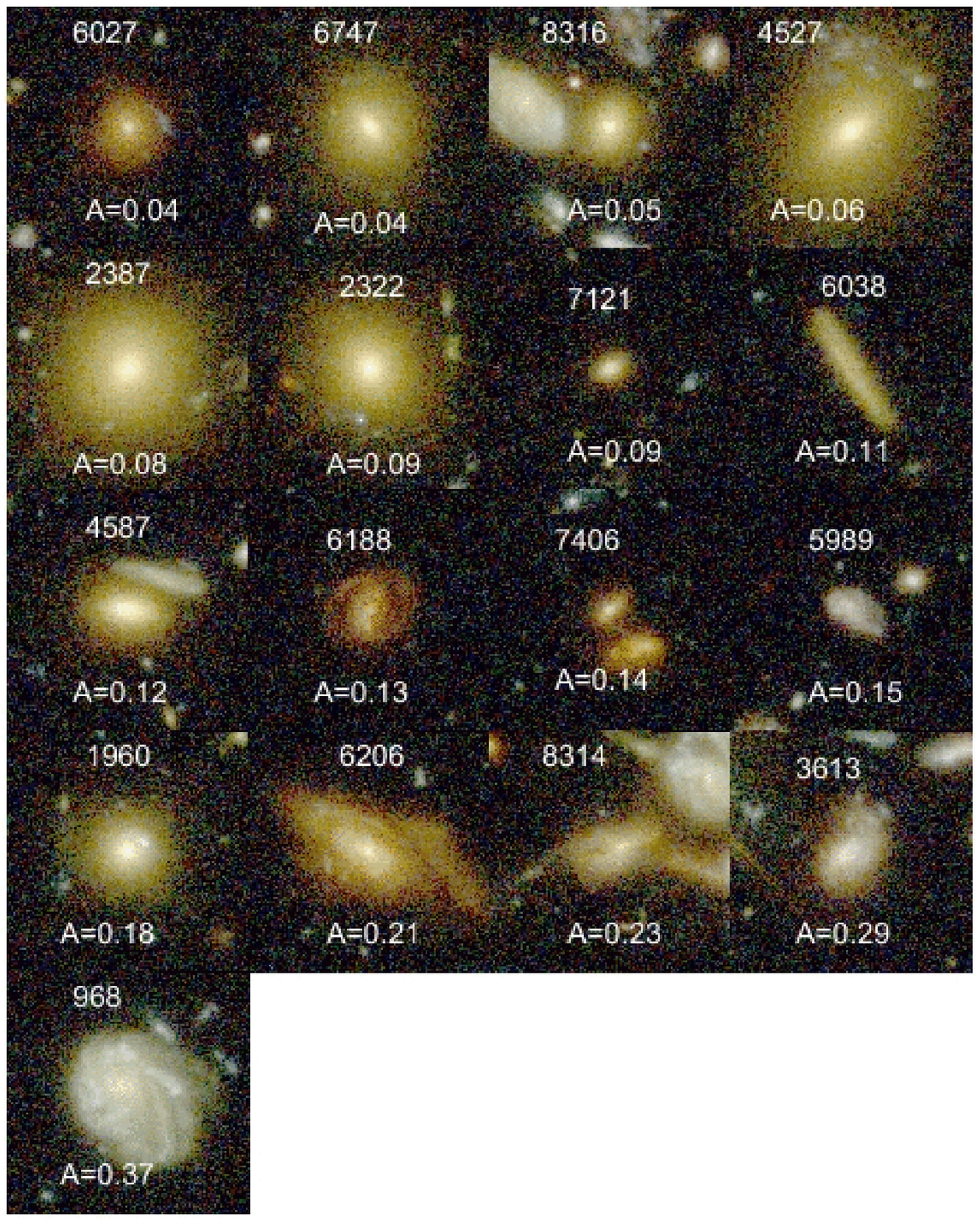,height=38pc}}
\caption{Galaxies in the Hubble Ultra Deep Field (UDF) as imaged through the ACS camera ordered
by how asymmetric these systems are.  These
are all galaxies with redshifts $0.5 < z < 1.2$ and with stellar masses M$_{*} > 10^{10}$ \solm.  
The ID is the number used in Conselice et al. (2008), and the $A$ value is the value of the asymmetry.
At these redshifts most of the massive galaxies can still be classified as being on the Hubble
sequence.}
\label{fig1}
\end{figure}

The critical nature of these deep fields is not simply that they are deeper 
than previous deep HST imaging, but that they are the fulcrum of large-scale 
efforts to obtain photometry at the faintest levels
possible at nearly all wavelengths.  This allowed redshifts to be measured
for most galaxies using photometry through so-called photometric redshifts (e.g.,
Dahlen et al. (2013) for a recent discussion of the use of this method). 
The availability of these redshifts allows us to measure evolution over broad
redshift ranges which was not possible before as spectroscopic
samples were few and far between, even with
the use of relatively large multi-slit spectrographs such as LRIS on Keck
(e.g., Cohen et al. 1996).

\begin{figure}%3	% Figure using psfig.sty
\centerline{\psfig{figure=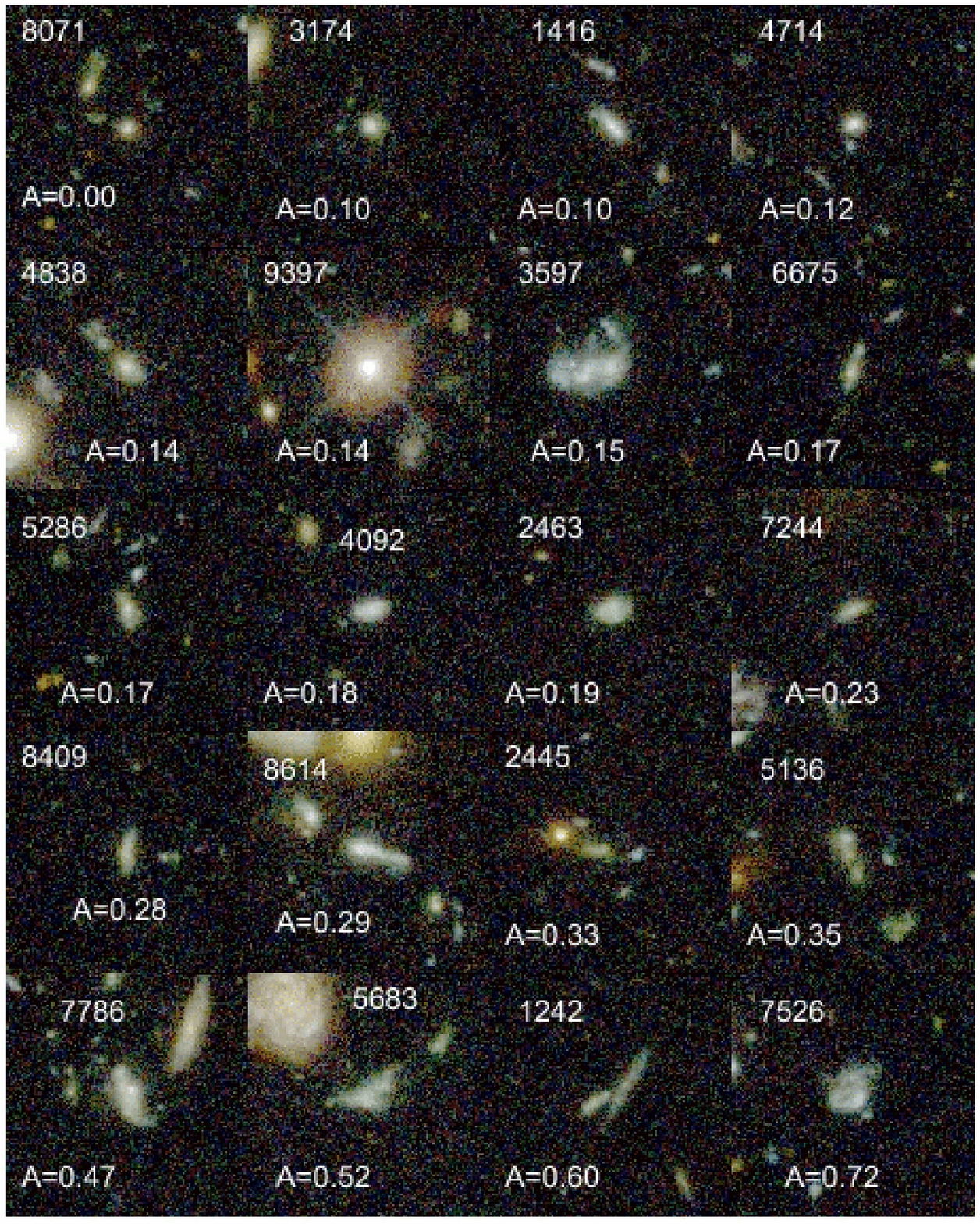,height=38pc}}
\caption{Similar to Figure~9, showing massive galaxies in the Hubble Ultra Deep Field (UDF) 
as imaged through the ACS camera ordered by the value of their asymmetries from most
symmetric to asymmetric.  Shown in this figure are systems with  stellar masses 
M$_{*} > 10^{10}$ \solm at redshifts $2.2 < z < 3$.
These galaxies are typically much smaller, bluer,
and have a higher asymmetry and inferred merger fraction than galaxies of comparable mass today
 (Conselice et al. 2008).}
\label{fig1}
\end{figure}

Early observations using a mixture of spectroscopic and photometric
redshifts showed that the Hubble sequence was
certainly not in place at high redshifts $z > 1$ (van den Bergh
1996) when examined using WFC2 data.  This was verified in the rest-frame optical 
after deep near-infrared NICMOS
observations of the HDF were taken in 1998 (Dickinson 2000).   A deeper
analysis of this showed that using the rest-frame optical structures of
galaxies, the Hubble sequence was nearly completely absent at $z > 2$, and
only at $z \sim 1.5$ did spirals and ellipticals become as common as peculiar
galaxies (Conselice et al. 2005; Buitrago et al. 2013; Figure~7).

The latest results on the evolution of visual morphology, as defined
solely by visual types, from the CANDELS survey, is shown in
Figure~8 (Mortlock et al. 2013) for systems with
stellar masses M$_{*} > 10^{10}$ \solm.    It must be stressed that this figure only 
shows the visual estimates of galaxy morphology.  The classification of a 
spiral, elliptical or peculiar does not imply that these galaxies have a 
certain star formation rate, color, mass or size. In fact what is often
seen is that these visual morphologies do not correlate well with other physical
properties (e.g., Conselice et al. 2011b; Mortlock et al. 2013).    What is also seen
is a stellar mass difference in the formation history of the Hubble
sequence, with the highest mass galaxies appearing to form into visual
ellipticals and disks before lower mass systems (Mortlock et al. 2013).

What Figures~7 and 8 show is that at $z > 2$ the dominant 
morphological type is peculiar, while disks and ellipticals become more common 
at lower redshifts (see also Conselice et al. 2005; Kriek et al. 2009; Delgado-Serrano 
et al. 2010; Mortlock et al. 2013).  
Figures~9 and 10 show images of the most massive galaxies at both $z < 1$ and 
$z > 2$ in the Hubble Ultra Deep Field, demonstrating the stark differences between
the two.   This review later examines in \S 4.3
what these peculiars are potentially evolving into at lower redshifts. There is
however a general trend such that peculiar systems, and galaxies with internal
features such as blobs, have a higher star formation rate than smoother
galaxies (e.g., Lee et al. 2013).    This shows that while the Hubble sequence
itself is not formed early in the history of the universe, there are trends
which suggest it is becoming established.

Another 
interesting aspect to examine is not only the fraction of different
galaxy types at various 
redshifts, but also their number density evolution as a function of galaxy type. 
We know from galaxy stellar mass measurements that the most massive systems are 
largely in place at $z \sim 1$ at the same number density as at $z \sim 1$ 
(e.g., Bundy et al. 2005; Mortlock et al. 2011).  It is the lower mass galaxies which 
typically increase the most in terms of the relative number density from the 
epoch $z \sim 0 - 2$.  In the nearby universe most of the massive galaxies are 
ellipticals, thus this implies that most of the evolution in morphology will 
also be within the lower mass galaxies, which is what is found.  In additional 
to a mass downsizing, we also find a morphological downsizing, such that 
morphological ellipticals are formed before the other galaxy types, most notably 
the spirals.  However, whether these galaxies are inherently similar to the 
ellipticals and 
disks today is a separate question.

\subsubsection{The Formation of Ellipticals and Disks - Bars, Bulges,  Disks, and Spiral Arms}

One of the major questions in galaxy evolution we would like to 
address is when modern spirals and ellipticals form.   This is related 
to the formation of
the Hubble sequence, but is a more detailed question, as what appears
as an elliptical/spiral may be quite different from systems classified
this way in the local universe.
We thus are interested in determining when systems with the same 
morphology to galaxies we see in the local universe (i.e., disks and
ellipticals) achieve a similar
physical state as measured through other properties.   While morphological 
fractions are similar at $z \sim 0.5-0.8$ as seen in the nearby universe, 
physical properties of galaxies with similar types - 
such as colors, star forming knots,  and tidal features, have not
reach the same level as locally.  It is only recently at $z < 0.3$
that the galaxy population appears in most ways similar to that
of today.  

This question can also be further divided into many sub-questions,
but for the purposes of this review I will examine the formation
history of galaxies identifiable as disks and ellipticals in their
gross morphology, as well as the formation of more detailed
features such as spiral arms,  bulges and bars.

One major issue is the bar fraction of galaxies and how it has 
evolved with time.  Early studies found that the bar fraction
evolves significantly (Abraham et al. 1999), while later studies find 
that bars were already in place up to $z \sim 1$ (Elmegreen,
Elmegreen \& Hirst 2004; Jogee et al.
2004).   Using the two degree area COSMOS HST survey Sheth et al. (2008)
find that the bar fraction increases from $z = 0.84$ to $z = 0.2$, 
from 20\% to $\sim$ 60\% of all disk galaxies.  Sheth
et al. (2008) however find that the bar fraction is roughly constant with redshift 
for the most massive and red disk galaxies. In fact most of the observed evolution  
occurs for the lower mass bluer disk galaxies.

The fraction of spiral galaxies with bars tells us when the disks in these 
galaxies become dynamically mature enough to form these structures. The 
fraction of bars also allows us to determine if bars have a role in the 
evolution of star formation, bulge
formation, and the triggering of AGN and the formation of supermassive
black holes.  The observational studies above all locate bars 
within disks either through changes in ellipticity and position angle in the
surface brightness profiles of galaxies, or through visual inspection.
It remains to be seen how the bar fraction holds when observed with a
near-infrared band such as within the CANDELS survey. 

A related issue is finding the onset of spiral structure, which
has remained a problem for
a variety of reasons, especially the difficulty of an unambiguous detection
due to resolution/depth issues.  Several papers published with the advent of ACS on
HST showed that there are many examples of disk-like morphological systems
at high redshifts. This includes the disks found by Labbe et al. (2003), the
Luminous Diffuse Objects of Conselice et al. (2004) and later systems identified by
Elmegreen et al. (2007) as clump-clusters.  More recently, Law et al. (2012) 
discovered a galaxy with a likely bona-fide spiral structure at
$z = 2.18$, which also contained a large internal velocity dispersion based
on IFU spectroscopy.  Spiral arms however appear very rare at high
redshift, and likely only a small number are formed before $z \sim 1$.
This is a major problem that needs more addressing and could be done
with recently available data.  However, there are many galaxies with clumpy
features that are potentially spiral arms and disks in formation which
we discuss in detail in \S 4.4.

Another feature of spirals and disks if the formation of bugles.  While the
traditional scenario is that bulges form in mergers (\S 4.3), recent results
suggest that bulges can form in multiple ways.  Psedo-bulges are different
from classical bulges in that they are likely formed through secular processes
within a disk.  This can be seen in the different correlations between bulge
properties and central massive black holes (e.g., Kormendy \& Ho 2013).

While morphology itself can often be ambiguous in terms of matching with
contemporary Hubble types, integral field unit (IFU) or long slit spectroscopy 
can remove some degeneracies.  In particular the use of integral field spectroscopy on
$z > 1$  systems reveals important clues about the nature of these high 
redshift galaxies (F\"orster Schreiber et al. 2009; Glazebrook et al. 2013).
Currently the most influential studies have been carried out with the SINFONI
IFU on the VLT, as well as some work done with OSIRIS on Keck (Law et al. 2012).

These surveys, most notably the SINS collaboration 
(e.g., F\"orster Schreiber et al. 2009) find an equal distribution of different
kinematic classes 
which are: rotationally dominated, mergers, and 
very compact galaxies with a high velocity dispersions (e.g., Buitrago
et al. 2013b for massive galaxies).  The true nature of 
these systems is not yet known, although in general it appears that galaxies
at $z > 2$
which are clumpy tend to have high velocity dispersions.  This perhaps reveals
the formation modes of disk galaxies at high redshift results in a high
velocity dispersion.  The future looks promising for combining larger
surveys of IFU measurements of distant galaxies with resolved morphologies to
decipher evolution.
Large surveys with for example KMOS on VLT and MOSFIRE on Keck
will revolutionize this area in the coming years.

\subsection{Size and Profile Shape Evolution}

One of the most important findings in galaxy evolution studies in the past
decade has been the discovery that distant galaxies are more
compact than systems of the same mass in the local universe
(e.g., Daddi et al. 2005; Trujillo et al. 2007; Buitrago
et al. 2008; van Dokkum et al. 2008, 2010; Weinzirl et al. 2011;
Baro et al. 2013; Williams et al. 2014).  

This change in sizes with time is now well characterized, and the 
evolution of galaxy sizes at a constant stellar mass selection
of M$_{*} > 10^{11}$ \solm
can be characterized by a power law of the form
$R_{\rm e} \sim \alpha (1+z)^{\beta}$.  The value of the power-law slope
changes with the galaxy surface brightness type, such that the disk-like galaxies with
S{\'e}rsic indices $n < 2.5$ evolve with $\beta = -0.82\pm0.03$, 
while spheroid-like galaxies with $n > 2.5$ have $\beta = -1.48\pm0.04$ 
(Figure~11). This demonstrates that
there is a faster evolution in measured sizes for spheroid-like galaxies, 
which therefore have a more effective increase in size over cosmic time than the 
disk-like objects.

\begin{figure}%3	% Figure using psfig.sty
\centerline{\psfig{figure=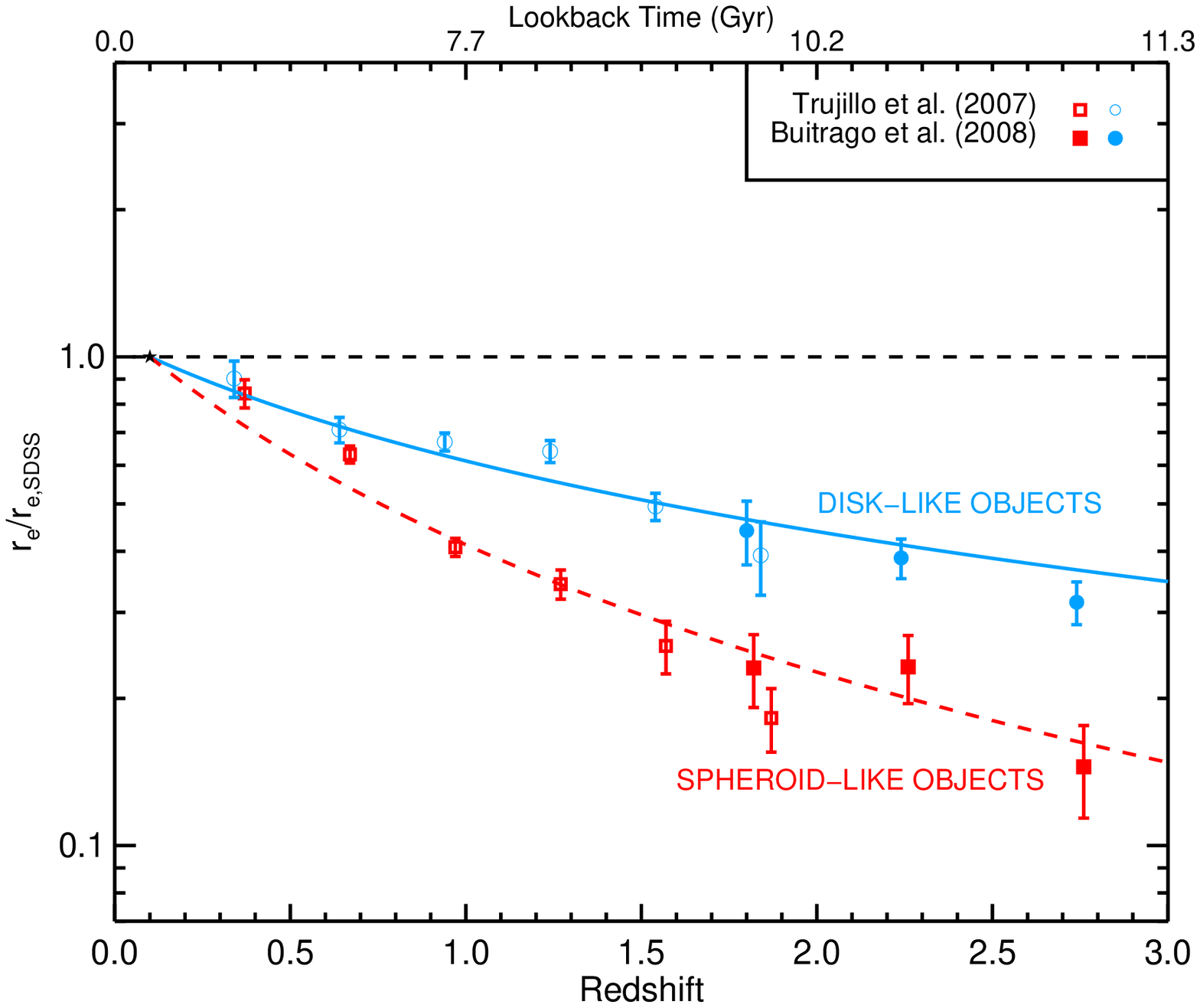,height=20pc}}
\caption{The average sizes of massive galaxies selected with M$_{*} > 10^{11}$
\solm as imaged in the POWIR (Conselice et al. 2007) $z < 2$ data and 
GNS $> 1.5$ images (Buitrago et al. 2008; Conselice et al. 2011).  The size
evolution is divided into galaxies with elliptical-like profiles, with
S{\'e}rsic indices $n > 2.5$, and disk-like profiles having $n < 2.5$.  The
measured effective radius, $r_{\rm e}$, is plotted as a function of the ratio
with the average size of galaxies at the same stellar mass measurements
with M$_{*} > 10^{11}$ \solm at $z = 0$ from Shen et al. (2003).  }
\label{fig1}
\end{figure}

This size evolution is such that the effective radii of massive galaxies
increases by up to a factor of five between $z = 3$ and today at the
same stellar mass (e.g.,
Buitrago et al. 2008; Cassata et al. 2013).  The
form of this evolution has been investigated to determine whether or
not the increase is due to the build up of the entire galaxy or just
the inner or outer parts.  The data to date show that galaxy growth
through sizes is occurring in its outer parts, with the central parts
in place at early times (e.g., Carrasco et al. 2010; van Dokkum et al. 2010).
This indicates that the build up of massive galaxies is an inside out
process, whereby the inner parts of massive galaxies are in place before
the outer parts with the same stellar mass density as today (e.g.,
Hopkins et al. 2009).

An alternative way to investigate this problem is to examine the
number of compact and ultra-compact galaxies at various redshifts.
There is some controversy over whether or not there exist in the local
universe compact galaxies with sizes similar to those seen at high
redshifts. However, what is clear is that the number densities
of these ultra-compact galaxies declines in relative abundance
very steeply at $z < 2$ (Cassata et al. 2013).

The processes responsible for this increase in sizes at lower
redshifts is not well understood, and is currently a source of
much debate.  The most popular explanation is
that this size increase is produced through minor mergers (e.g., Bluck
et al. 2012; McLure et al. 2013), although other ideas such as AGN performing
work on gas is another idea (e.g., Bluck et al. 2011).  However, the outer parts of
nearby massive galaxies are too old to have been formed in relatively
recent star formation, and the star formation observed at high redshift
is not sufficient to produce the observed increase in sizes
(Ownsworth et al. 2012).

The major idea for the physical mechanism producing galaxy size 
evolution is through dry minor mergers, as major mergers are not
able to produce the observation of increasing size without
significantly increasing mass (e.g., Khochfar \& Silk 2006;
Naab et al. 2009; Bluck et al. 2012; Oser et al. 2012; Shankar et al. 2013).
There is currently some controversy  over whether or not the
observed minor merger rate is high enough to provide this increase
in sizes, with the most massive galaxies with M$_{*} > 10^{11}$ \solm appearing
to have enough minor mergers (e.g., Kaviraj et al. 2009) to
produce this size evolution  (Bluck et al. 2012), but this may not be the
case for lower mass systems (e.g., Newman et al. 2012).  It does
appear however that minor mergers are a significant mechanism for producing
low levels of star formation in early-types at $z \sim 0.8$, as well as 
for adding significant
amount of stellar mass to these galaxies (Kaviraj et al. 2009, 2011).
 One of the major issues
 is determining not only the number of
minor dry mergers, but also the time-scale for these mergers (\S 3.4)
which more simulations would help understand.

Along with the evolution of galaxy sizes, there is also a significant
evolution in the underlying structures of galaxies at higher
redshifts. One of the cleanest ways to see this is through the
evolution of the S{\'e}rsic parameter, $n$ (Figure~7).  When examining
the evolution of derived values of $n$ as a function of redshift for both
a stellar mass and at a constant number density selection, it is apparent that
galaxies have lower $n$ values at higher redshifts for the same
selection (e.g., Buitrago et al. 2013).    this has been
interpreted by some to imply that these galaxies are more 'disk-like'
at high redshifts (Bruce et al. 2012), although the morphologies of these systems by
visual inspection, and their internal structures and colors, are not
similar to modern disks (e.g., Conselice et al. 2011; Mortlock et al. 2013).  
It appears
that these disk-like galaxies, while having light profiles similar to 
modern disks, are much smaller, have a higher stellar mass, and are often
undergoing intense star formation with peculiar morphologies, making
them un-disk-like in all other regards. They indeed are likely a type
of galaxy with no local counterpart.

\subsection{The Merger History}

One of the primary, if not the primary use of galaxy structure at high redshift 
at the time of writing, is using it to measure the merger history of galaxies.  
This is a major issue in extragalactic astronomy, as merging is not only a 
method for galaxies to form, but is also a potential way in which black 
holes, star formation and other
internal features of galaxies are assembled.   Merging is also one of
the key predictions of the Cold Dark Matter model which is the dominant idea
for how galaxy evolution occurs through cosmic time (e.g., White \& Rees 1978;
Blumenthal et al. 1984; Cole et al. 2000).

The merger history of galaxies was first measured through galaxies
in pairs - systems of at least two galaxies near enough to each other to
merge in a reasonable amount of time
(typically 20 or 30 kpc), and in the case of kinematic pairs with a 
low velocity difference of around 200 \kms 
(e.g., Patton et al. 1997, 2002;  Le F{\'e}vre et al. 2000; Lopez-Sanjuan 
et al. 2011, 2012). 
Simulation results show that most pairs of galaxies selected in this
way will eventually merge, and within a time-scale comparable to dynamical
friction (e.g., Moreno et al 2013). 
Using pairs to find galaxies which are merging is still a large industry, and 
interested readers are referred to the latest papers in this field 
(e.g., Lopez-Sanjuan et al. 2011, 2012; Tasca et al. 2013).
Pairs of galaxies also provide a check on the methodology we use to 
find major mergers, as it is an independent measure of this formation mode.
However the merger fraction measured through pairs is more
statistical in nature than structure, and does not reveal for certain whether a 
galaxy is a merger or not.

A detailed CAS structural study of starburst, ULIRGs, and other merging 
galaxies shows that another route for measuring the merger history is through 
the use of the asymmetry index, whereby the most asymmetric galaxies are ones 
involved in mergers (Conselice 1997; Conselice et al. 2000a,b; Conselice 2003;
\S 3.2).   The merger history is also measured through other parameters, 
whereby mergers occupy a unique parameter space (e.g., Lotz et al. 2004; 2008; Freeman
et al. 2013).  

While there is no perfect 1:1 relation between parameter space definitions of
mergers with all
mergers identifiable by eye, we still find that about half of all identifiable 
mergers fall
within specific regions of parameter space, and that there is a low contamination 
from other galaxy types (\S 3.2).     Thus, galaxy merger fractions are an
observation which has to be interpreted carefully with the use of galaxy
merger models.

\subsubsection{The Merger Fraction Evolution}

The merger fraction history is the basic observable which reveals how mergers
are changing and evolving through cosmic time.  It is measured at high 
redshift through the criteria described in \S 3.2, and the resulting merger
fractions are shown in Figure~12 (e.g., Conselice et al. 2008, 2009;
Bluck et al. 2012).
This figure shows that the inferred merger fraction increases with redshift.
This increase is typically fit as a power-law of the form:

\begin{equation}
f_{\rm m} = f_{0} \times (1+z)^{m}
\end{equation}

\noindent where $f_{0}$ is the merger fraction at $z = 0$ and $m$ is the power-law 
index for measuring mergers.  In general, the higher the value of $m$ the more 
steeply the merger history increases at higher redshifts.  An alternative
parameterization of the merger history is given by a combined power-law
exponential (e.g., Conselice et al. 2008) whose form is:

\begin{equation}
f_{\rm m} = \alpha \times (1+z)^{m} {\rm exp} [\beta(1+z)].
\end{equation}

\noindent The local $z =0$ merger fraction in this formalism is
given by $f_{\rm m}(0) = \alpha\, {\rm exp}(\beta)$, and the merger 
peak is located at $z_{\rm peak} = -(1+m/\beta)$.  This
form of the merger fraction evolution appears to fit lower mass 
galaxy merger fractions better than a single power-law up to $z \sim 3$, 
as this allows for a peak at a value of the mergers and a decline at
higher redshifts.   In fact, 
only the highest mass galaxies with M$_{*} > 10^{10}$ \solm appear to increase
up to $z \sim 3$, while lower mass galaxies have a merger
peak around $z \sim 1.5-2.5$, and declining at higher 
redshifts (e.g., Conselice et al. 2008).

While it initially appeared that there were significant differences in merger
histories between different studies (e.g., Lin et al. 2004) it is now clear that
these are due to several effects.  The first is that the value of the power-law
index $m$ can vary significantly just due to the value of the anchor redshift
at $z = 0$.  Secondly, when comparisons are done between galaxies selected in
the same way (stellar mass or absolute magnitude), and correct time-scales are
used for different techniques (\S 3.4) then merger rates agree within the
uncertainties (e.g., Conselice et al. 2009; Lotz et al. 2011).

The merger history tends to peak at $z \sim 2.5$ for massive galaxies with
M$_{*} > 10^{10}$ \solm at values of $f_{\rm m} \sim 0.3-0.4$ and
decline at lower redshifts.  The values for $m$ found in the literature can vary significantly,
but most of this is due to different selections, different redshift ranges used, and the use of various
values for the local merger fraction.  The first studies using pairs find a very steep increase
up to $z \sim 1$, with $m = 2.8\pm0.9$ for a luminosity selected sample up to $z \sim 0.4$ (Patton
et al. 1997).  The value of $m$ was later found by Patton et al. (2002) to be $m = 2.3\pm0.7$ within
the CNOC2 redshift survey up to $z \sim 0.55$.  Le F{\'e}vre et al. (2000) measure the merger fraction 
using 285 galaxies in the CFRS and LDSS surveys up to $z \sim 1$
finding a power-law index of $m = 3.2\pm0.6$, although this
lowers to $m = 2.7\pm0.6$ after considering selection effects.    Other studies
have found similar values, with $m = 1.5\pm0.7$ for brighter
galaxies in the VVDS (de Ravel et al. 2009), and $m = 3.1\pm0.1$
for pairs up to $z \sim 1$ in the COSMOS field (Kartaltepe 
et al. 2007).   These are however
all relatively nearby galaxies at $z < 1$.  For higher redshifts, Bluck et al. (2009)
find a power-law of $m = 3.0\pm0.4$ for galaxies with M$_{*} > 10^{11}$ \solm using
pairs from the GOODS NICMOS Survey.

Within morphological studies, Conselice et al. (2003) find a high $m$ 
index of $m \sim 4$ for massive galaxies with M$_{*} > 10^{10}$ \solm, 
with $m$ values around $m \sim 1-2$ for  lower mass galaxies.  This is 
similar to what is found by Le F{\'e}vre et al. (2000) when
examining the merger history of visually disturbed galaxies in Hubble imaging.   
This was confirmed by Conselice et al. (2008) using the same
methods, but on the Hubble Ultra Deep Field data.  
Conselice et al. (2003) also show how the merger fraction slope $m$ 
can vary significantly depending on what redshift limit is used, and 
whether stellar mass or luminosity cuts are applied to the selected sample.   
In a detailed study of
$z < 1$ galaxies from the COSMOS survey, Conselice et al. (2009) find an index
of $m = 2.3\pm0.4$ for galaxies with stellar masses of M$_{*} > 10^{10}$ \solm. On 
the other hand, Lotz et al. (2008) find that the merger fraction does not evolve
significantly with redshift between $z \sim 0.2 - 1.2$, with a weak increase 
in the merger fraction with $m = 0.23\pm1.03$ using the Gini/M$_{20}$ methods.
Later however it was shown that Gini/M$_{20}$ is very sensitive to 
minor mergers, and once these effects are considered the fitted parameter is
$m = 2$ up to $z \sim 1.2$ (Lotz et al. 2011).  Other methods of measuring
the merger or interaction rate includes looking for
ringed galaxies (D'Onghia et al. 2008), a method which derives a merger
fraction index with $m \sim 3$.

\subsubsection{Galaxy Merger Rate Evolution}

The merger fraction is simply just an observational quantity,
as it is the fraction of galaxies in a sample which have merged with another galaxy.
This merger fraction
can effectively be anything from zero to near unity depending on the mass ratio
and time-scales of interest.   Merger fractions must therefore be carefully
interpreted.  The physical quantity we are interested in is
the merger rate, which gives the number of mergers occurring per unit time and
in some cases per unit volume per unit time.

Bluck et al. (2012), Lotz et al. (2008) and Conselice (2006b) show that 
CAS is only sensitive to major mergers of mass ratios of 4:1 and greater, and
has a particular time-scale of about $\sim 0.5$ Gyr associated with the 
merger sensitivity (\S 3.4).
This is very similar to the merger parameter sensitivity when using pairs of 
galaxies with mass ratios of 4:1 or greater (i.e., major pairs).  Thus 
by examining the merger fraction with CAS we are likely tracing the same systems 
as measured through galaxies in pairs.

The merger fraction is converted into a galaxy merger rate through the use of 
the merger time-scale, $\tau_{\rm m}$ which can be derived through simulations 
(Lotz et al. 2010a) (see \S 3.4), or through the decline in the merger fraction 
at lower 
redshifts (Conselice 2009). The result of both approaches is that the CAS merger 
time-scale is around 0.5 Gyr for a galaxy to remain 'peculiar' in parameter space, 
with the merger time-scale in Gini/M$_{20}$ at a similar level but which is 
more sensitive to minor mergers (\S 3.4). 

Using the merger time scale, the merger rate per galaxy is thereby defined by the 
parameter $\Gamma$ (e.g. Bluck et al. 2009; Conselice et al. 2009),

\begin{equation}
\Gamma (M_{*}, z) = \frac{\tau_{\rm m}}{f_{\rm gm}}.
\end{equation}
 
\noindent The value of $\Gamma$ is in units of Gyr giving
the average amount of time between mergers for a galaxy within the 
given selection property, typically 
stellar mass,  and within a given redshift range.    Note that
within the definition of $\Gamma$ the galaxy merger fraction is
used rather than the merger fraction (\S 3.2).   The value of
$\Gamma$ as a function of redshifts for CAS merger measures
and pair selection mergers sample is shown in Figure~13.

Using the evolution of the galaxy merger rate, per galaxy, $\Gamma$, the 
the number of mergers that occur between various redshifts is calculated by 
integrating the inverse of $\Gamma$ over redshift,

\begin{equation}
N_{\rm merg} = \int^{t_2}_{t_1} \Gamma^{-1} dt = \int^{z_2}_{z_1} \Gamma^{-1} \frac{t_{H}}{(1+z)} \frac{dz}{E(z)},
\end{equation}

\noindent  The result of this is studied in detail in Conselice et al. (2009) and 
Bluck et al. (2009) with initial results using simulations discussed in
Conselice (2006b).  The result of these calculations show that the number of 
major mergers a galaxy undergoes between $z = 3$ and $z = 1$ is 4.3$\pm0.8$ major 
mergers at $z < 3$ (Conselice et al. 2008) for galaxies selected with
stellar masses M$_{*} > 10^{10}$ \solm.  There also appears to be a limited 
number of mergers at higher redshifts $z > 3$ (Conselice \& Arnold 2009).

The $\Gamma$ definition
of the merger rate is per galaxy and thus does not take into account the
overall merger rate within the universe.  Thus
ultimately we are interested in the galaxy merger rate, $\Re({\rm M_{*}}, z)$

\begin{equation}
\Re(z,M_{*}) = f_{\rm gm} \tau_{\rm m}^{-1} n_{m},
\end{equation}

\noindent where $n_{m}$ is the number density of galaxies at a given redshift.
While this is the ultimate quantity in galaxy merger studies, it is difficult
to measure, and the number density, $n_{\rm m}$, has its own associated uncertainties 
(e.g., Mortlock
et al. 2011).  In Figure~13 we plot the merger rate for galaxies using
the CAS systems for pairs and mergers using the number densities of
galaxies from Mortlock et al. (2011).

While the number of mergers is an interesting and fundamental quantity, we 
are ultimately interested in the amount of stellar mass added from mergers to galaxies 
over time.  The total amount of stellar 
mass accreted into a galaxy is calculated as a double integral over the 
redshift range of interest ($z_1$ to $z_2$ or look-back times $t_1$ and $t_2$), 
and over the stellar masses range in which is being probed ($M_1$ to $M_2$), and
can be expressed as,

\begin{equation}
{\rm M_{*, M}} = \int_{t_{1}}^{t_{2}} \int_{M_{1}}^{M_{2}} M_{*} \times \frac{f'_{m}(z,M_{*})}{\tau_{\rm m}(M_{*})} dM_{*} dz,
\end{equation}

\noindent where $\tau_{\rm m}$(M$_{*}$) is the merger time-scale, which 
depends on the
stellar mass of the merging pair (Bluck et al. 2012).  From this we
calculate the total 
integration of the amount of mass assembled through merging for galaxies
with stellar masses M$_{*} > 10^{11}$ \solm.  For a $z < 3$
mass complete sample at this limit, Conselice et al. (2013) find a value
${\rm M}_{\rm *,M}/{\rm M_{*}(0)} = 0.56\pm0.15$, where M$_{*}(0)$ is 
the initial average stellar mass at $z \sim 3$, and ${\rm M}_{\rm *,M}$ is the 
average amount of stellar mass accreted in mergers at $z = 1 - 3$.   This ratio 
is the fractional amount of stellar mass added both due to major and minor mergers
for systems with stellar mass ratios down to 1:100 for an average massive
galaxy after following a merger adjusted constant co-moving 
density (Conselice et al. 2013).   By observing the number of mergers
we can get an idea of how much gas is added to galaxies through the merger
process. Comparing this to the star formation rate there is a large deficiency,
which must be accounted for by gas accreted from the intergalactic medium.  
Using this number and the observed star formation rate within these galaxies, the 
gas accretion rate from
the intergalactic medium is calculated at around 100 \solm year$^{-1}$, adding
roughly the same amount of stellar mass to these galaxies as mergers
(Conselice et al. 2013).  

%which for thesensitive down to M$_{*} = 10^{9.5}$ \sol,

\begin{figure}%3	% Figure using psfig.sty
\centerline{\psfig{figure=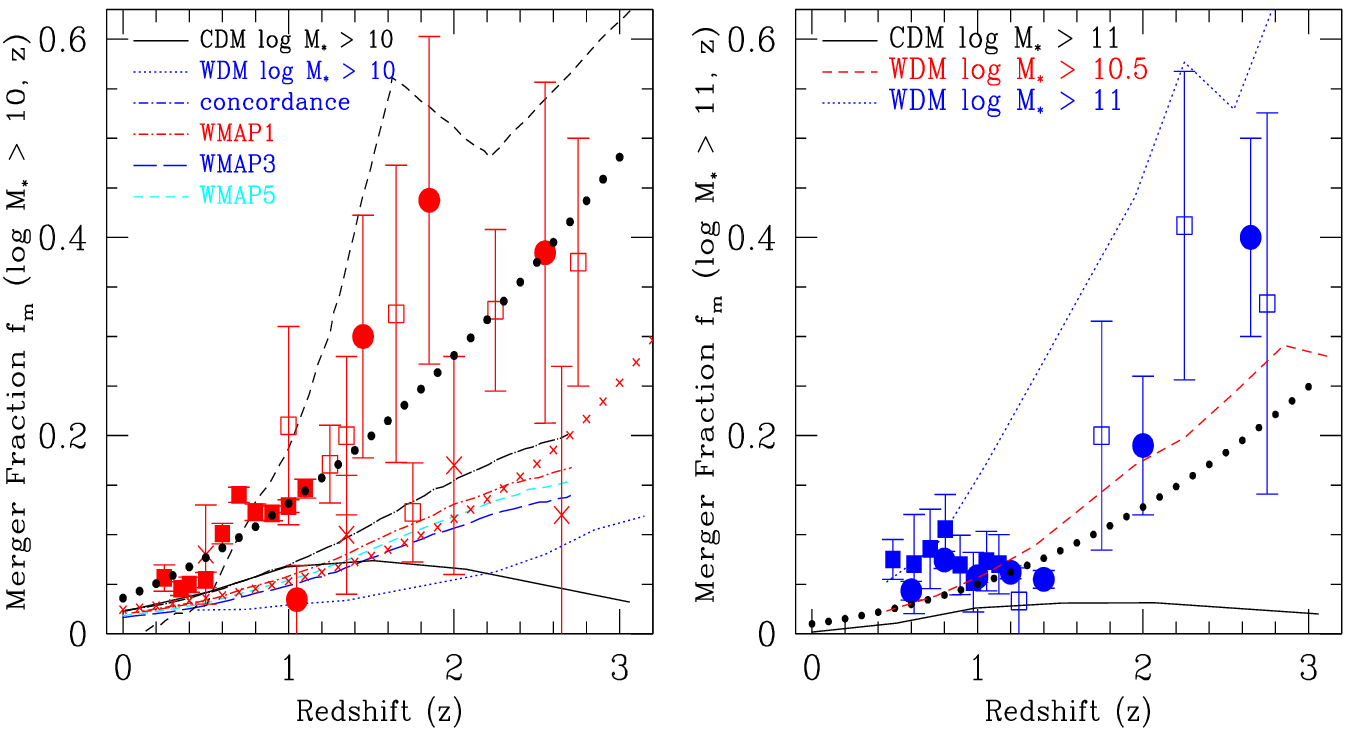,height=16pc}}
\caption{A compilation of the  merger fraction history
for galaxies selected with stellar masses M$_{*} > 10^{10}$ \solm and
M$_{*} > 10^{11}$ \solm.  The points shown here on the
left panel are for  M$_{*} > 10^{10}$ \solm
galaxies, including results from Conselice et al. (2003) at $z > 1$ in
the HDF (solid circles); Conselice et al. (2009) (solid boxes at
$z < 1.2$); and Mortlock et al. (2013) (open boxes at $z > 1$).   Also 
shown on the left panel are pair merger fractions at separations of $< 30$ kpc: (Man et al.
2012; crosses at $z > 1$); Lopez-Sanjuan et al. (2010) (large open boxes).
The right panel shows the merger history for M$_{*} > 10^{11}$ \solm
systems including results from Conselice
et al. (2009) (solid boxes at $z < 1.2$; Bluck et al. (2009, 2012)
(solid circles at $z > 0.5$); and Mortlock et al. (2013) (open boxes
at $z > 1$).    In both panels the line with the dark solid circles is the best fit 
relation for a merger fraction parameterization as $\sim (1+z)^{m}$.  
The blue dotted line on the left, and the red and cyan dashed lines on the
right show the Warm Dark Matter predictions, and the solid black line the
Cold Dark Matter predictions using semi-analytical models.    For 
the M$_{*} > 10^{10}$ \solm panel we also show the merger fraction
calculation from abundance matching using various WMAP and concordance
cosmologies (shown on panel), as well as abundance matching from Stewart
et al. (2008) (red crosses). }
\label{fig1}
\end{figure}

\begin{figure}%3	% Figure using psfig.sty
\centerline{\psfig{figure=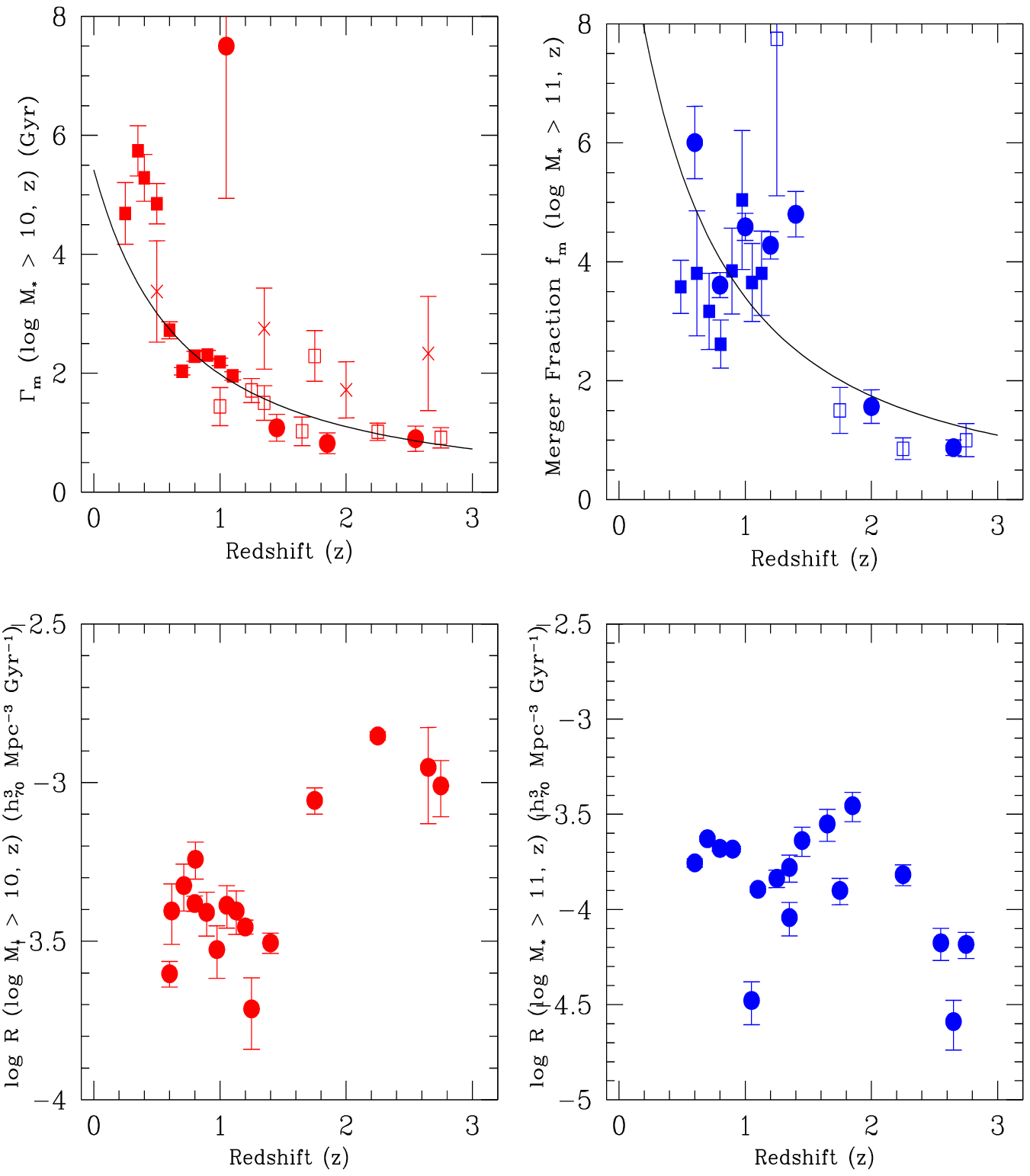,height=30pc}}
\caption{The merger rate for galaxies plotted in two
different ways.  The
upper panels plot the merger rate per galaxies, $\Gamma (z)$
at stellar mass limits of M$_{*} > 10^{10}$ \solm (left) and
M$_{*} > 10^{11}$ \solm (right).   The units of
$\Gamma$ are Gyr, and represent the average time between
mergers for a typical galaxy at the given mass limit. The ultimate realization of
the merger rate is shown at these two stellar mass limits
in the bottom row, as the number of mergers occurring per Gyr
per co-moving Mpc$^{3}$.    The point types
in the merger rate per galaxy has the same meaning as
the merger fraction plot in Figure~12.}
\label{fig1}
\end{figure}

\subsection{Resolved Morphological Formation Histories}

Independent of the formation of the gross, or bulk, galaxy structure 
is the formation of their stars, and how this correlates with structure and its
assembly.
This is effectively done by examining the spectral energy distributions
of individual components in galaxies (i.e., disks or bulges) and/or a 
newer technique of examining the SEDs of individual pixels.  With resolved
imaging it is therefore possible to determine to some extend the star
formation history of individual resolution elements (e.g., Abraham
et al. 1999; Zibetti et al. 2009) as well as the stellar
mass within these (e.g., Lanyon-Foster et al. 2007).

\begin{figure}%3	% Figure using psfig.sty
\centerline{\psfig{figure=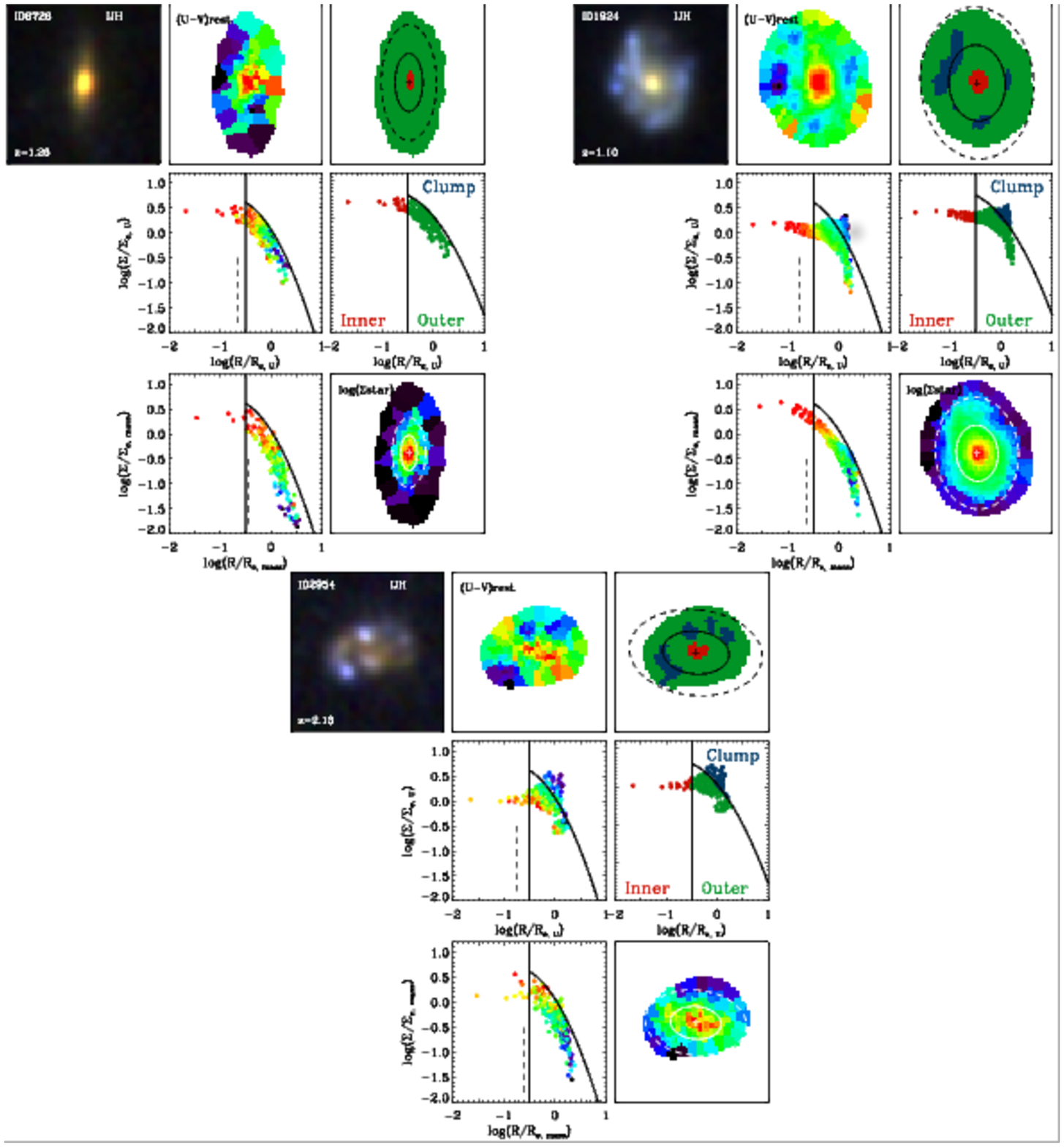,height=34pc}}
\caption{Internal pixel structures and profiles for galaxies at $z \sim 1-2$ with and
without star formation.  The examples are ID6726 which is a non-star forming
galaxy, and ID12924 and ID1954 which contain clumpy star forming knots throughout their
structures.  These are representative examples of commonly found galaxies
at high-redshifts, typically around $z \sim 2$ as seen in deep WFC3 images from surveys
such as CANDELS.  For each of these galaxies the top rows demonstrate the image in the
IJH HST bands, the rest-frame (U-V) color map, and a map which indicates whether pixels
belong to the inner (colored red), outer (green) or clump region (dark blue)
for each galaxy.  The solid black ellipses 
show the area which contains half of the rest-frame U-band light for each system. The
second row for each galaxy shows the U-band light profile which is color coded by
the pixel type labeled in the $(U-B)$ color map and on the right by the location
of the various pixels.  The lines shown in the middle panels show the separation
between the different spatial distributions.   The bottom panel shows the distribution
in terms of stellar mass.  The vertical dashed line shows the  resolution limit of WFC3. 
The final panel shows the stellar mass map for each system, with the inner solid and outer 
dashed lines showing the location of $R_{\rm e, mass}$ and $2 R_{\rm e, mass}$. From Wuyts
et al. (2012) and courtesy of Stijn Wuyts.}
\label{fig1}
\end{figure}

Early results using small samples of few dozen galaxies at $z \sim 1$, Abraham et al. (1999)
demonstrated that many morphologically selected ellipticals in their sample show a 
diversity in their star formation histories, with recent bursts of
star formation commonly seen.  Abraham et al. 
(1999) also showed that bulges in these systems almost always have ages older than their 
disks.    This was followed up with a study of 79 field 
spheroids by Menanteau et al. (2001) who find that a third of their 
morphologically regular
systems have recent star formation in their centers, with so-called 'blue cores'.
It is possible that some of these systems are forming the pseudo-bulges that
we see in the nearby universe.

There are other studies of the bulges of distant galaxies, including Hathi 
et al. (2009) who find that bulges at $0.8 < z < 1.3$ have ages of a 1-2 Gyr, 
with stellar
masses up to 10$^{10}$ \solm in the Hubble Ultra Deep Field.   They also
find that late type bulges are younger than early types, a finding which
also exists in the local universe.  However, classical bulges up to 
$z \sim 1$ with de Vaucouleur profiles are found to have old stellar populations
similar to giant ellipticals at the same epoch (Koo et al. 2005).

Another potentially powerful approach towards understanding the formation history
of galaxies is to examine their light distribution on a pixel by pixel
basis. The idea here is that each pixel or resolution element is independent of
others and each has their own SED which can be fit by stellar population
analysis methods (e.g., Bothun 1986; 
Abraham et al. 1999; Lanyon-Foster et al. 2007).  Local galaxies have different 
pixel color magnitude diagrams, depending on morphological type,
with early types having a much narrow locus of points than disk galaxies.

Galaxies also often look more symmetrical in stellar mass maps than in 
light (Lanyon-Foster et al. 2012; Wuyts et al. 2012' Figure~14). 
Star forming galaxies in the CANDELS survey were recently studied in a pixel by
pixel approach in  Wuyts et al. (2012)
who examined 323 systems at $0.5 < z < 1.5$ and a further 326  higher redshift
systems at $1.5 < z < 2.5$.  This study uses the optical ACS and near infrared 
WFC3 filters to construct SEDs for all pixels in these galaxies.  Wuyts
et al. (2012) find that the nuclei of star forming
galaxies are redder and have older ages than their outer parts. Clumps in
these galaxies also generally occupy a smaller fraction of the total mass
than the total light, demonstrating that the large clumps seen in distant
star forming galaxies are mostly star forming regions.   Clumps are also
found to be central, off-central, or outer, and these clumps may play a 
role in forming the bulges of these galaxies through a secular process
(Bournaud et al. 2007; Elmegreen et al. 2008; Genzel et al. 2008) (\S 5).
This shows that there are methods beyond hierarchical clustering for the
formation of structures within galaxies.

\subsection{High Redshift AGN/Starbursts and Star Formation Quenching}

One of the major ideas behind how mergers drive galaxy formation is that
when galaxies merge, gas clouds collide, triggering star formation.  At
the same time 
gas is driven to the centers of galaxies, producing active galactic
nuclei (AGN) (e.g., Hopkins et al. 2008).  Observationally we also know
that galaxies which are concentrated, with for example a high
S{\'e}rsic profile, are more likely to be quenched (e.g., Bell
et al. 2012; Fan et al. 2013).  There is therefore a strong
theoretical reason to expect that structure, as influenced by mergers,
should correlate with galaxy formation.  There is also strong evidence
that the star formation in galaxies is regulated by structure.  How
these relate however is not yet clear.

Observational it has proven thus far difficult to correlate the
presence of mergers, through either pairs or
 the presence of a distorted or peculiar structures, with the
presence of AGN or a star formation excess at high redshift.  
Recent studies such as 
Cotini et al. (2013) use CAS indices to show that the fraction of
nearby galaxies with AGN are roughly five times more likely than
a control sample to have a distorted structure.   However, studies
at higher redshifts generally do not find that more asymmetric or
merging galaxies have a higher AGN fraction (e.g., Grogin et al.
2005; Gabor et al. 2009; Kocevski et al. 2012).
This is also found in the deepest ACS Hubble imaging, although it
does appear that AGN are found in more concentrated galaxies at 
higher redshifts (Grogin et al. 2005). There is also a lack of
a higher fraction of peculiar/merging galaxies found in the
GEMS/STAGES HST survey by Bohm et al. (2013) for AGN with luminosities of
L$_{\rm X} < 10^{44}$ erg s$^{-1}$. This may be an indication that 
Type-1 AGN are more likely to be found in early-type massive systems.
It remains to be seen if obscured AGN, the Type-2s, are found in
more merging systems and thereby represent an earlier phase
of the merger.

Another major issue in which morphology and structure play an important 
role is understanding the origin of dusty star forming galaxies, 
the so-called ULIRGs or sub-mm galaxies.  Sub-mm
galaxies are those which appear very bright at sub-mm wavelengths,
typically at 850 $\mu$m, and are at high-redshift at $z > 1$.  
Morphological analyses of these sub-mm galaxies
showed early on that they are involved in merger activity, and have
peculiar morphologies (e.g., Chapman et al. 2003; Conselice et al.
2003b).    Many of these galaxies have structures and morphologies
consist with being involved in mergers (e.g., Ricciardelli et al. 2011).
Kartaltepe et al. (2012) find that a sample of ULIRGs selected from the
CANDELS fields are more likely than a field galaxy sample to be involved
in galaxy interactions and mergers (72$^{+5}_{-7}$\% vs. 32$\pm3$\%).
However, Swinbank et al. (2010) argue using a large sample of
sub-mm galaxies that their morphologies are not significantly 
different from other star forming field galaxies at similar redshifts.
Swinbank et al. (2010) also find that their  sub-mm galaxies have light profiles 
more similar to
early-types than disks.   Although, Targett et al. (2011) find that 
sub-mm galaxies have disk-like profiles, and conclude that these systems are
more like forming disks than spheroids forming in mergers.  
This suggests that the morphology of 
sub-mm galaxies is still open for debate with results that appear to
be conflicting.  More detailed work, likely with adaptive optics in the
K-band to avoid issues with dust and morphological k-corrections, are
needed to make further progress.

Furthermore, as mentioned earlier there is a strong observed correlation between 
galaxies which have steep surface brightness profiles, with $n > 2.5$, and
the quenching of star formation (e.g., Bell 2012).  This shows that
galaxy structure either produces a change in the galaxy history, or more
likely is
a symptom of the effects which produces this quenching.  In general, there are
a few ways to quench the star formation seen in galaxies
at high redshift.  Some of these are environmental, such as ram-pressure
stripping and strangulation (e.g., Peng et al. 2010).  However, 
these processes will be ineffective for
the bulk of field galaxies which we study in this review.  What is more
likely driving the quenching of typical field galaxies is merger
driven, or driven by the stellar/halo mass in some form of feedback
process
(e.g., Peng et al. 2010; Peng et al. 2012; Lilly et al. 2013; Carollo
et al. 2013).   The most likely candidates are mergers that either heats
gas or removes most of it in giant starbursts, preventing further star formation.  
The other idea for
this feedback is that it is the result of AGN.  The idea here is
that the ongoing star formation is truncated by the existing gas in a galaxy
being heated or removed by an active AGN (e.g., Croton et al. 2006). 
It is therefore likely that a few critical processes
are ongoing to produce the Hubble sequence, and specifically the
red, passive and concentrated high mass systems.

\subsection{The $z \sim 3-6$ Frontier}

By far the bulk of what we know of galaxy structural evolution is at 
$z < 3$. The reason for this is simply because this is
the limit where we can observe the rest-frame optical using observations
in the near-infrared from the Hubble Space Telescope, which is only effective
at imaging at filters bluer than the H-band (often H$_{160}$ with WFC3 and
NICMOS).    However, there are some observations of galaxy structure at even higher
redshifts that provides some information about the formation of these galaxies.
To date most of these observations are done in a red filter using the ACS
camera on Hubble, either the I$_{814}$ or z$_{850}$ band.
It must be remembered that these systems are being observed in the rest-frame
ultraviolet, and thus their morphologies will be dominated by young stars.

There have only been a few major studies that focus on the structures of these
ultra-high redshift galaxies.  Ferguson et al. (2004) study the sizes and
the axis ratios of Lyman-break galaxies (LBGs) galaxies up to $z \sim 6$ 
in the GOODS fields, finding
that galaxies are smaller and more 'disk-like' in their axis ratios at higher
redshifts.  This was also shown in an extensive study of 4700
LBGs by Ravindranath et al. (2006) who find that 40\% have
exponential light profiles, 30\% have de Vaucouleurs profiles, and the
remaining 30\% have multiple cores.  The ellipticity distribution of these
LBGs shows that these systems are skewed towards high values with
$\epsilon > 0.5$, which cannot be explained by viewing disks and spheroids
at various angles (Ravindranath et al. 2006).  This is either
an indication that these systems are mergers, or that star formation
is distributed in the outer parts of these galaxies creating these elliptical
structures.

For ultra-high redshift galaxies, Conselice \& Arnold (2009) examine the visual
morphologies and pair fractions of Lyman-break galaxies in the Hubble Ultra Deep Field
ACS Filters from $z \sim 4 - 6$.  These ACS data on the UDF still provide the 
highest resolution and deepest imaging
of the most distant galaxies.  Conselice \& Arnold (2009) find that the
fraction of $z \sim 3-6$ Lyman-break galaxies which are peculiar in appearance is
roughly constant at $\sim 30$\% throughout this redshift range.  Conselice \&
Arnold furthermore demonstrate that many of the LBGs at these redshifts have
tidal like features -- fans, shells, etc. that resemble merger signatures 
seen at lower
redshifts.    The derived merger fraction from LBG pairs also agrees with the
merger fraction based on CAS and visual estimates (Conselice \& Arnold 2009;
Cooke et al. 2010).  

It is perhaps surprisingly easier
to identify galaxies in pairs at these redshifts than at lower redshifts,
as one can use the drop-out band
to remove contamination, and thus ensure that two galaxies close by in the sky 
are at least at a similar redshift.  This results in a smaller
correction needed to calculate merger fractions, and thus the merger
fraction in principle can be measured more accurately (Conselice \& Arnold 2009).
Jiang et al. (2013) similarly examine the rest-frame UV morphologies of 
51 Lyman$-\alpha$ galaxies and 16 Lyman-break galaxies and find a merger 
fraction for the brightest galaxies of around 50\%, and otherwise a diversity 
in morphology.  

Most recently, using WFC3 data from the UDF Oesch et al. (2010) show 
that $z \sim 7-8$ galaxies are very compact
with a typical size of 0.7$\pm 0.3$ kpc with little size evolution down 
to $z \sim 6$. There is more development down to $z \sim 4$ with
the observation of more extensive wings of light at these lower redshifts,
and a corresponding increase in sizes (e.g., Ferguson et al. 2004) following
a similar power-law with redshift, as is found for size
evolution between $z \sim 3$ and $z \sim 0$ 
(Buitrago et al. 2008; Mosleh et al 2012).

\subsection{Role of Environment  in Structure Formation}

Galaxy morphology is well known to
correlate strongly in the local universe with environment
(e.g., Dressler 1984; Postman et al. 2005).   It is also clear that there is
a strong relationship between morphology and stellar mass,
such that the most massive galaxies tend to have elliptical
morphologies and lack star formation.  Combining this 
with the morphology-redshift relation shows that the
structure of a galaxy depends upon its mass, local environment,
as well as time.  Which of these is the leading  cause for
producing galaxy evolution is an active area of study.

The problem of galaxy morphology as a function of density is
a large area of research and is outside the immediate
scope of this article.  However, it is relevant to discuss some of
the major findings, and how they relate to the evolution of galaxy
structure with time.  The major effect of morphology is that the
type of galaxy, either elliptical or spiral, depends to a large
degree in the nearby universe on the local density of that particular
galaxy's environment.  This relation is such that the higher
the density of the local environment, the more likely a galaxy
will be early-type and non-star forming (e.g., Dressler 1984;
Gomez et al. 2003; Blanton \& Moustakas 2009).  Disk properties 
are also highly environmentally
driven, with few classical bulge or elliptical systems in low density
environments (Kormendy et al. 2010).

It is also the case that more massive galaxies are more likely
to be early type.  The question is which relationship 
is more fundamental, and relates to the old issue of ''nature vs. nurture'' for
galaxy formation.  The structures and morphologies of galaxies
can help address this problem, especially by examining the limited
number of observations we have of galaxy structure in high redshift
overdensities, or (proto-)clusters.  

Observations of overdensities at high redshifts are just starting in
earnest, but already provide some clues to this.  Very massive clusters 
at high redshifts,
 up to $z \sim 1.2$, contain a similar pattern of morphologies 
and densities as local galaxies, such that the denser areas contain a 
higher fraction of systems which are elliptical.  
This tends to break down for the limited number of cluster candidates found at higher
redshifts, where the galaxy population is irregular and peculiar, as is
found for the general high redshift galaxy population (e.g., Papovich et
al. 2012).

Detailed studies are however possible up to $z \sim 1$ both by using
field galaxies of various local environments, as well as looking at
the morphological and structural distributions of galaxies within
rich clusters at various redshifts.  For field galaxies, Tasca
et al. (2009) examine the morphology-density relationship for 100,000
galaxies in the COSMOS survey.  They find that the
morphology-density relation changes slightly with redshift,
becoming flatter at higher-z (e.g., Gr{\"u}tzbauch et al. 2011a,b).  
Above a stellar mass of about
10$^{10.6}$ \solm the morphologies of galaxies appear to become
more dominated by the stellar mass as the critical factor rather
than density (e.g., Gr{\"u}tzbauch et al. 2011a).  The situation 
in rich clusters at $z \sim 1$, and the formation of
S0s, is such that the trend with environment is not as steep as is
found at $z \sim 0$. This suggests that S0s are not entirely formed
yet in these distant clusters, however the elliptical population does
seem to be in place compared with the population at $z \sim 0$.

Another structural feature that can be investigated is the size evolution
and how it varies with environment. The limited number of investigations of 
this have found that
galaxies at $z > 1$ in higher density environments show signs of a more rapid
increase of galaxy size with redshift in comparison to the field (e.g.,
Cooper et al. 2012; Lani et al. 2013).  This is one indication
whereby a dense environment can facilitate a more rapid evolution in
galaxies, although 
it is a slight effect that needs further confirmation.

\section{Comparisons to Theory}

Galaxy morphology and structure allows a new way to compare with cosmologically
based galaxy formation
models,  as well as those which include extensive physics
such as star formation, AGN feedback and supernova in more detailed hydrodynamical
models.  This review only briefly discusses this large topic and how it
relates to galaxy structure.  For a more detailed recent review on the theory of
galaxy formation from a theoretical prospective see Silk
\& Mamon (2012). 

Galaxy formation models were first developed to explain the 
structures of galaxies, namely
the bulge/disk/halo trichotomy, and the ages of the stars in these components
(Eggen et al. 1962).  The default initial assumption in the first galaxy
formation models was that galaxies formed
like stars in a relatively rapid collapse.  In the 1980s the first 
computer simulations  of structure formation  showed 
that a universe dominated by Cold Dark Matter (CDM) matched observations of galaxy 
clustering on large scales (Davis et al. 1985), and that within this 
framework galaxy assembly should be hierarchical (Blumenthal et al. 1984),
yet this is a fundamental prediction which is just now starting to be tested 
with only a
few papers  comparing the observations to the theoretical predictions
(e.g., Bertone \& Conselice 2009; Jogee et al. 2009; Hopkins et al. 2010;
Lotz et al. 2011).

The situation today is that there are many simulations that are used to 
predict properties of the galaxy population, and how it evolves through
time. These models are largely successful when predicting basic properties
of nearby galaxies, such as their luminosities, masses, colors and star formation rates, 
as well as  scaling relationships of galaxies. However problems still exist in
predicting the abundances of low and high mass galaxies
(e.g., Conselice et al. 2007; Guo et al. 2011).    
Within galaxy formation models there are very famous problems such as the 
satellite and the CDM dark matter profile, but there are also significant
issues when examining 
how the evolution of galaxies occurs, and trying to match this with the theory.
Another major problem is that there are several large disk galaxies without 
significant bulges in the nearby universe that are not predicted in CDM
(e.g., Kormendy et al. 2010).

One of the ways to further test these models is to investigate 
how well CDM models can reproduce the formation history of galaxies 
as seen through the  merging process using the so-called semi-analytical method 
(e.g., Bower et al. 2006; Guo et al. 2011).
We show this comparison with the measured merger fractions in Figure~12
at two different stellar mass ranges of M$_{*} > 10^{10}$ \solm and
M$_{*} > 10^{11}$ \solm.  Plotted as the thin solid black line towards
the lower part of each diagram is the prediction for the major merger
fraction for galaxies from the Millennium simulation (Bertone \& Conselice
2009).  Also shown on these figures as the dotted blue line is the
same predictions for major mergers for Warm Dark Matter models
(e.g., Menci et al. 2012), which do a better job than CDM in matching
the observed data.  However, CDM better matches if minor mergers are
taken into account, although the comparison merger fraction is
only for major mergers based on the methodology used (Conselice
2003a; Lotz et al. 2010a). 

Other recent attempts to predict the merger history of observed galaxies
include the abundance matching technique (e.g., Stewart et al. 2008) 
where observed galaxies are matched to halos in models through their
comparative abundance levels.  Hopkins et al. (2010) predict based on 
this abundance matching the merger rate and fraction for galaxies. The 
result of this is
show in Figure~12 for galaxies between M$_{*} = 10^{10-11}$ \solm.  While
the merger fractions from Hopkins et al. (2010) are higher than those from
the CDM models, they are still lower than the observations (see also
Jogee et al. 2009 and Lotz et al. 2011 for further discussions).  Similar
results from Stewart et al. (2008) are also shown in Figure~12, who find
results similar to Hopkins et al. (2010).

Finally, as a contract to these  Maller
et al. (2006) present cosmological hydrodynamical simulation results for 
similar mass galaxies of a few times 10$^{10}$ \solm, and find the highest
merger fraction predictions of any simulation result (Figure~12).    This
shows that the predictions for merger histories are not correct or
consistent with each other, and that
more simulation work should be focused on this critical aspect of the
galaxy population.  This is an area where future work is certainly
needed.

There are several other types of simulations in which galaxy structure
and morphology can be directly compared with observations of galaxies
through cosmic time.    Perhaps the most direct of these is to compare the
properties and structural features of distant galaxies to 
hydrodynamical models of galaxy formation.   Some of this work
for galaxy mergers is discussed in \S 3.4.  Early work in this area showed
that the components of galaxies - namely bulges and disks were the result
of accretion events (e.g., Steinmetz \& Navarro 2002) and argued
from their simulations that the Hubble type of a galaxy is not stable
for long periods of cosmic time.  Governato et al. (2007) show that disk
galaxies can be simulated which have properties that match the morphological properties and
kinematics of nearby disks, although this simulation is not in a cosmological
context.   Overall however it is very difficult to predict the formation of
galaxy morphology in simulations, and in a real sense this will be
the ultimate test of galaxy formation models in the future.

Also, as discussed in \S 4.6 one of the most commonly seen properties in high
redshift star forming galaxies is that they often contain large clumps of
star formation within their disks.  A major question is how these clumps form,
evolve, and how they may play a role in the formation of other galaxy components
such as bulges and AGN.  Bournaud et al. (2013) examined this problem computationally
to determine how clumps with stellar masses of M$_{*} = 10^{8-9}$ \solm evolve
in gaseous disks.  The major question is whether these clumps dissipate within 50 Myr
or so, the dynamical time-scale of the clumps, or if they regenerate and
survive.  Bournaud et al. (2013) find that
these clumps can last around 300 Myr through acquiring new gas from its disk, although
some mass is lost through tidal effects.  This is enough time to migrate towards
the center of the galaxy which can fuel the AGN or merge to form a bulge.  This
shows that these clumps may provide a significant route for galaxies to form.
Thus, we have evidence for both inside-out and outside-in formation occurring in
the galaxy population.  What remains to be seen is whether one of these mechanisms
is dominant, and the relative role of both in forming galaxies.

\section{Summary and the Future}

I present here a review of galaxy structure and morphology studies in the
galaxy population through cosmic time from $z = 8$ until today.  The
approach taken in this review is largely observational with a limited
amount of interpretation, although I do show where galaxy structure and
morphology can test galaxy formation and even cosmological models in
a new, largely unexplored way.  

As of January 2014, the major conclusions concerning galaxy structure
and its evolution, as discussed in this article, can be summarized as:

\vspace{0.2cm}

\noindent I. Galaxy structure and morphology is the longest studied observational
feature of galaxies.  In this review galaxy morphology is the apparent classification
based on visual inspection, while structure is a way to quantify the light
distributions in galaxies.   In many ways morphology is still a descriptive 
science, and visual
efforts continue to provide useful information in the form of large-scale 
projects
to classify many galaxies, as in the Galaxy Zoo effort.
The Hubble sequence has and likely will remain the major paradigm in which
we consider galaxy morphology, although this system does not 'work' at
high redshifts where most galaxies cannot be classified into a single
Hubble type (\S 4.1).

\vspace{0.2cm}

\noindent II. Using the Hubble scheme the evolution of three broad classes
of galaxies are now classified accurately out to $z \sim 3$ -- namely
ellipticals, spirals and peculiars.  The relative abundance of these
galaxies has been measured as a function of redshift out to these
early times. What we find is that the peculiar galaxies dominate the
galaxy population at $z \sim 2.5 - 3$, with a relative fraction of
at least 70\%.  Galaxies which are elliptical and spiral like
in appearance (but not necessarily in physical properties, see \S 4.1)
become progressively more common at lower redshifts.  The number densities of
these two normal galaxies together equals that of the peculiars by $z \sim 1.4$
(Mortlock et al. 2013).

\vspace{0.2cm}

\noindent III.  Since galaxy morphology by visual estimates is limited
in its ability to derive the physics behind galaxy formation and by its
nature is not quantitative, the use of parametric (\S 2.2) and non-parametric
(\S 2.3) methods are essential for deriving in a quantitative way how galaxies
are evolving.   These quantitative indices also correlate to some degree
with the present and past star formation history and properties of
a galaxy.  More work is needed to establish these relations with
more certainty, but it appears that the S{\'e}rsic index and concentration
correlate with the scale or mass of a galaxy, the clumpiness index
with the star formation, and the asymmetry parameter with ongoing
merging activity (\S 3).  

\vspace{0.2cm}

\noindent IV. The merger history is now know from applying structural analyses 
to galaxy images
in deep Hubble Space Telescope surveys such as the Hubble Deep Field (\S 4).
The result of this is that the galaxy merger fraction increases with redshift
at all stellar mass and luminosity selections.  This increase can be fit well
by a power-law $(1+z)^{m}$ up to $z \sim 3$, although at higher redshifts the
structurally derived merger fraction may plateau (Conselice \& Arnold 2009).
Using numerical/hydrodynamical simulations the time-scales for these mergers
can be calculated, and thus merger fractions can be converted into merger
rates (\S 4.3.2). The merger rate allows for the calculation of the number of
mergers galaxies at various masses undergo, as well as the amount of stellar mass
which is added to galaxies due to the merger process.  The result of
this is that it appears that up to half of the stellar mass in modern massive
galaxies were formed in mergers between $1 < z < 3$, although at $z < 1$ dry
mergers are likely more responsible for the further formation of these
galaxies.

\vspace{0.2cm}

\noindent V. The resolved structures of galaxies also allows us to measure
the internal features of galaxies and how they are assembling. There is some
controversy over the formation history bulges, disks and bars, although many of
these are likely formed by secular processes produced internally by disk 
dynamical evolution.  This is an area
where significant progress could be made in the next few years.  The most
up to date results suggest that the bar fraction for spiral galaxies
at $z < 1$ depends upon the stellar mass of the galaxy. The most massive
galaxies have a similar bar fraction at $z \sim 0.8$ as they do today,
yet lower mass and bluer disk galaxies have a significantly lower bar 
fraction than similarly low mass nearby disks. This mirrors the evolution of
the Hubble sequence itself where more massive galaxies settle into normal ellipticals
and disks before lower mass galaxies.  Spiral structure is a difficult problem
and while some examples exist at high redshift, even at $z > 2$, the general
onset of when disks form spirals is almost totally unconstrained by observations.

\vspace{0.2cm}

\noindent VI.  Resolved imaging also permits us to measure the spectral energy
distributions and colors, of galaxy components and individual pixels of different
galaxies. This is another area where more work needs to be performed, but it appears
that bulges of spirals tend to be older than their disks at high redshift, but
there are examples of many ellipticals which have blue cores and central
star formation (\S 4.1.1).  Pixel-pixel analyses show that galaxies have a mixed
star formation history, and that the inner parts of galaxies are often older
than their outer parts.  Pixel studies also show that the clumps seen in distant
star forming galaxies are composed of young stellar populations, and thus must
have recently formed or regenerate themselves.

\vspace{0.2cm}

\noindent VII.  Perhaps the most popular (at present) problem in galaxy structural
evolution is the apparent compactness in size of galaxies at high redshifts. The
observations show that massive galaxies at $z > 1$ have sizes which are a factor of
2-5 smaller than similar massive galaxies in today's universe.  This result has
been studied in many different ways, and the sizes of a stellar mass selected sample
of galaxies increases gradually as a function of redshift with a power-law
increase $\sim (1+z)^{\beta}$, where $\beta$ varies from $-0.8$ to $-1.5$ depending
upon whether the selected samples are disk-like or elliptical-like (\S 4.2).  Results
to date suggest that these galaxies are building up their outer parts over time
to become larger systems, rather than adding mass to their centers.  This process
is unlikely driven by star formation, and theory suggests that this formation
is produced by minor merger events (\S 4.2).

In summary, we have learned much about galaxy morphology and structure over the
past 15 years.  There are however many open questions still remaining on all
aspects of using structure to determine evolution. More work needs to be done
in tying galaxy structure to underlying physics, both through empirical
work and in simulations.  Furthermore, the time-scales for structural features
such as mergers and large clump survival are critical to better understand.  
Broad morphological features
will remain important over the next decades as telescopes such as JWST, Euclid,
LSST, the Dark Energy Survey, amongst others, will all resolve many more galaxies
than we can currently study, and at higher redshifts.  This opens up entirely new 
possibilities, and the 
with careful thoughtful planning a new revolution in galaxy structure may be
upon us soon.

A review such as this is written with help from many people. In particular I
thank  Alice Mortlock, Asa Bluck, Fernando Buitrago, 
Jamie Ownsworth, and Ken Duncan for illuminating conversations and collaboration
on these topics over the past few years. 
I also thank Jennifer Lotz, Fernando Buitrago, Stijn Wuyts, Alice Mortlock for kindly
providing figures.  I personally thank the STFC and the Leverhulme Trust 
in the UK, as well as the NSF and NASA in the USA  for financial support.

%%% Numbered Literature Cited
\section{NUMBERED LITERATURE CITED}

\end{document}